\documentclass[11pt]{article}

\usepackage[final]{acl}

\usepackage{times}
\usepackage{latexsym}

\usepackage[T1]{fontenc}

\usepackage[utf8]{inputenc}

\usepackage{microtype}

\usepackage{inconsolata}

\usepackage{graphicx}
\usepackage{booktabs}
\usepackage{tabularx}
\usepackage{multirow}
\usepackage{makecell}
\usepackage{enumitem}
\usepackage{soul}
\usepackage[table]{xcolor}

\usepackage[most]{tcolorbox}
\usepackage{xcolor}

\usepackage{xcolor}
\definecolor{method_green}{HTML}{2ca02c}
\definecolor{method_orange}{HTML}{d55e00}

\newtcolorbox{aclprompt}[1]{
    enhanced,
    colback=blue!3!white,        
    colframe=blue!50!black,      
    boxrule=0pt,                 
    leftrule=3.5pt,            
    arc=0.5mm,                   
    width=\textwidth,          
    left=5mm, right=5mm, top=7mm, bottom=5mm,
    fontupper=\footnotesize\ttfamily\linespread{1.05}\selectfont,
    before=\vspace{12pt},
    after=\vspace{12pt},
    attach boxed title to top left={xshift=0mm, yshift=-3mm},
    boxed title style={colback=blue!50!black, colframe=blue!50!black, sharp corners, boxrule=0pt},
    title=#1,
    fonttitle=\bfseries\sffamily\scriptsize,
}

\newcommand{\var}[1]{\textbf{\textcolor{blue!70!black}{\{#1\}}}}

\definecolor{lightbrown}{RGB}{245, 222, 179} 
\definecolor{mybrandcolor}{RGB}{228, 213, 199} 
\definecolor{tablegray}{gray}{0.95}        

\newcommand{\sumdataset}{ClassEval-RE}
\newcommand{\deftectdataset}{DebugBench-RE}
\newcommand{\methodname}{VERA}

\title{Beyond Output Correctness: Benchmarking and Evaluating \\ Large Language Model Reasoning in Coding Tasks}

\author{
  \textbf{Yuangang Li\textsuperscript{1,*,†}}, 
  \textbf{Justin Tian Jin Chen\textsuperscript{1,*}}, 
  \textbf{Ethan Yu\textsuperscript{1}}, 
  \\\textbf{David Hong\textsuperscript{1}}, 
  \textbf{Iftekhar Ahmed\textsuperscript{1}} \\
  \textsuperscript{1}University of California, Irvine \\
  \texttt{\{yuanganl, chenjt3, ethay12, dbhong, iftekha\}@uci.edu} \\
  \textsuperscript{†}Corresponding author.
}

\begin{document}
\maketitle

\begin{abstract}
Large language models (LLMs) increasingly rely on explicit reasoning to solve coding tasks, yet evaluating the quality of this reasoning remains challenging. Existing reasoning evaluators are not designed for coding, and current benchmarks focus primarily on code generation, leaving other coding tasks largely unexplored. We introduce CodeRQ-Bench, the first benchmark for evaluating LLM reasoning quality across three coding task categories: generation, summarization, and classification. Using this benchmark, we analyze 1,069 mismatch cases from existing evaluators, identify five recurring limitations, and derive four design insights for reasoning evaluation in coding tasks. Guided by these insights, we propose VERA, a two-stage evaluator that combines evidence-grounded verification with ambiguity-aware score correction. Experiments on CodeRQ-Bench show that VERA consistently outperforms strong baselines across four datasets, improving AUCROC by up to 0.26 and AUPRC by up to 0.21. We release CodeRQ-Bench at \url{https://github.com/MrLYG/CodeRQ-Bench}, supporting future investigations.


\end{abstract}

\section{Introduction}

Large Language models (LLMs) are widely used for coding tasks~\cite{roziere2023code,hou2024large,zhang2024surveylargelanguagemodels,sun2025source,bouzenia2025repairagent}. With the rise of reasoning-enhanced models such as OpenAI’s o-series~\cite{jaech2024openai,el2025competitive}, DeepSeek-R1~\cite{guo2025deepseek}, and Chain-of-Thought (CoT) prompting~\cite{wei2022chain} and its variants~\cite{yao2023tree,besta2024graph,a2026asymmetric}, eliciting step-by-step reasoning has become a key strategy for improving performance on coding tasks.

\begin{figure}[t!]
\centering
\includegraphics[width=0.96\linewidth]{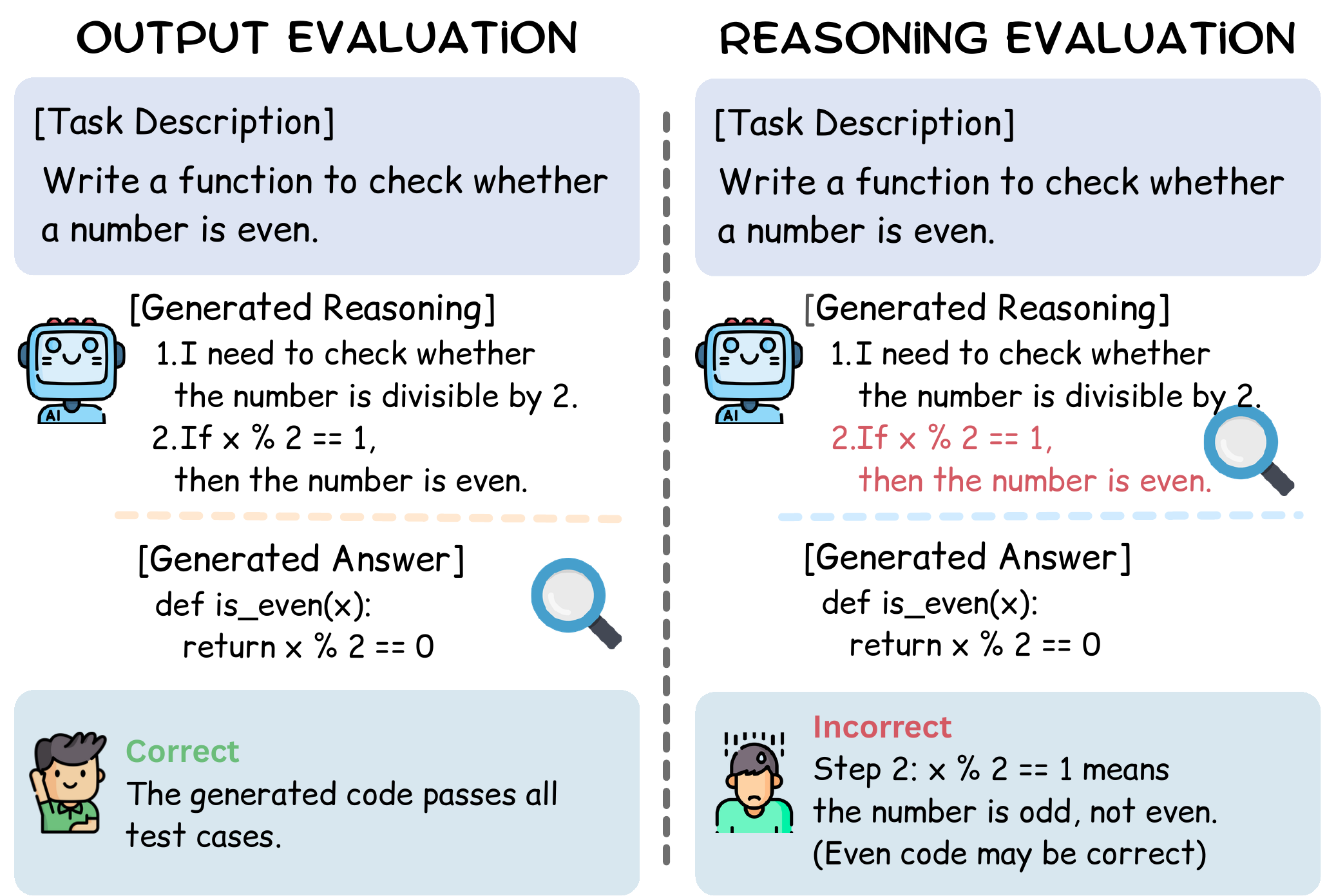}
\caption{Output-based vs. Reasoning-based evaluation in a code generation task.}
\label{fig:fig1}
\vspace{-0.27in}
\end{figure}

However, the quality of such reasoning is often unreliable~\cite{turpin2023language,lanham2023measuring,tanneru2024hardness,li2025mitigating}, and flawed reasoning can degrade the final output~\cite{zhang2025they}. Despite this, existing evaluation primarily focuses on final output correctness~\cite{sun2026agents}. Benchmarks such as SWE-bench~\cite{jimenez2024swebench} and HumanEval~\cite{chen2021evaluating} assess task success using outcome-based metrics (e.g., test passing, pass@k), leaving the reasoning process largely unexamined. As a result, LLMs may produce correct code while relying on faulty intermediate reasoning~\cite{zhang2025they} (Fig.~\ref{fig:fig1} shows an example). Ignoring the reasoning process limits our understanding of model behavior and hinders progress toward reliable reasoning in coding tasks~\cite{gao2026clusters}.

Recent work has begun to move beyond output correctness and evaluate the reasoning processes generated by LLMs. RECEVAL~\cite{prasad2023receval} assesses reasoning along correctness and informativeness, SOCREVAL~\cite{he2024socreval} uses Socratic-style prompting to evaluate reasoning, and CaSE~\cite{do2025defines} applies step-level causal scoring to assess relevance and coherence. However, these methods were developed and validated on general NLP tasks, such as arithmetic reasoning and commonsense question answering. Coding tasks differ from NLP tasks substantially, as they involve program structure, semantics, and execution behavior~\cite{liu2024codemind,yang2025code}. Therefore, it is necessary to evaluate these methods in the coding domain to assess their effectiveness.

Another key challenge is the lack of comprehensive benchmarks for evaluating reasoning quality in coding tasks. Existing efforts focus mainly on code generation, with limited coverage of other task categories. Coding tasks are commonly grouped by output modality into generation, summarization, and classification~\cite{lu2021codexglue,zhang2024surveylargelanguagemodels}. These categories require different forms of reasoning~\cite{gu2024cruxeval,liu2024codemind,yuan2025understanding,xie2025core,gao2026orchestration}: generation involves constructive planning, summarization requires semantic abstraction and compression, and classification relies on diagnostic deduction. Restricting evaluation to generation therefore provides only a partial view of evaluator performance, obscuring task-specific limitations and hindering the development of more reliable reasoning evaluators.

To address these limitations, 
\textit{(1)} we introduce \textbf{Code} \textbf{R}easoning \textbf{Q}uality \textbf{Bench}mark (CodeRQ-Bench), to our knowledge the first benchmark for evaluating the quality of LLM-generated reasoning across three coding task categories. 
Each instance is annotated for reasoning quality through a three-expert consensus-based annotation process, following established practices in dataset annotation and annotator agreement assessment~\cite{pustejovsky2012natural,teruel2018increasing,oortwijn2021interrater}.
\textit{(2)} Built on CodeRQ-Bench, we show that existing reasoning evaluators have five limitations on coding tasks. We analyze 1,069 cases where evaluators misjudge reasoning correctness, identify five recurring limitations, and derive four key insights that highlight new directions for reasoning-quality evaluation in coding tasks (see \S\ref{sec:analyzing_limitations}).
\textit{(3)} Motivated by these findings, we propose \textbf{V}erified \textbf{E}valuation of \textbf{R}easoning with \textbf{A}mbiguity-awareness (\methodname{}), a two-stage evaluator that first verifies reasoning correctness against external evidence and then calibrates scores based on task ambiguity and how well the reasoning addresses it. On CodeRQ-Bench, \methodname{} achieves state-of-the-art performance, improving over existing baselines by up to 0.26 in AUCROC and 0.21 in AUPRC.

\vspace{-0.1in}
\section{Related Work}

\paragraph{Outcome-Based Evaluation of Coding Tasks.}

Existing evaluation of coding tasks is largely outcome-based, relying on metrics such as execution correctness, exact match, and classification accuracy across tasks like code generation, repair, summarization, and bug detection~\cite{chen2021evaluating,du2023classeval,liu2023your,tian2024debugbench,jimenez2024swebench}. While effective for measuring task success, these metrics cannot assess the reliability of the reasoning process: correct outputs may arise from flawed reasoning, and vice versa~\cite{zhang2025they}. Recent benchmarks such as CRUXEval~\cite{gu2024cruxeval}, CoRe~\cite{xie2025core}, CodeMind, and EquiBench~\cite{liu2024codemind,wei2025equibench} evaluate models on reasoning-intensive coding tasks but still focus on task performance as the primary metric. In contrast, our work evaluates the reasoning chain itself, enabling direct assessment of reasoning reliability and more fine-grained diagnosis of reasoning failures.
\vspace{-0.1in}
\paragraph{Existing Reasoning Evaluation Methods.}

Early work evaluates model-generated reasoning by comparing it with human-written explanations using similarity-based metrics~\cite{clinciu2021study,welleck2022naturalprover}. ROSCOE extends this approach by assessing reasoning along multiple dimensions, including factuality, coherence, and redundancy~\cite{golovneva2022roscoe}. However, reference-based methods rely on predefined reasoning chains, limiting their ability to capture the diversity of valid reasoning paths.

Recent work moves toward reference-free reasoning evaluation. ReCEval~\cite{prasad2023receval} evaluates reasoning in terms of correctness and informativeness; SocREval~\cite{he2024socreval} uses Socratic-style prompting to guide LLMs through probing questions before producing an evaluation; and CaSE~\cite{do2025defines} applies causal stepwise scoring to assess relevance and coherence. However, these methods were developed for general reasoning tasks, such as arithmetic and commonsense question answering, rather than coding tasks. In coding settings, reasoning quality depends not only on linguistic coherence but also on code semantics, execution behavior, and technical constraints. We address this gap by introducing a coding-oriented benchmark and evaluator.

\section{CodeRQ-Bench: Code Reasoning Quality Benchmark}

\subsection{Preliminaries and Problem Definition}

Code Reasoning Quality (CodeRQ) evaluation focuses on assessing the correctness of LLM-generated reasoning for solving a coding task.
In CodeRQ evaluation, the input consists of a coding-task description $q$, an LLM-generated reasoning chain $C=(s_1,\ldots,s_n)$, and the corresponding final output $o$. 
Given $(q,C,o)$, the goal is to return a score in $[0,1]$ indicating how likely the reasoning process is correct, where larger values indicate higher estimated reasoning quality.

\vspace{-0.1in}
\subsection{Dataset Construction}
To date, the only reasoning-annotated datasets for coding tasks are CoderEval-\textbf{R}easoning \textbf{E}valuation (CoderEval-RE) and SWE-bench-\textbf{R}easoning \textbf{E}valuation (SWEbench-RE), both introduced by Zhang et al.~\cite{zhang2025they}. We adopt the DeepSeek-R1\cite{guo2025deepseek}-generated traces provided in that work (see Appx.~\ref{appx:dataset_details} for details).
A comprehensive benchmark for CodeRQ evaluation should cover the three categories of coding tasks under the output-modality taxonomy: generation, summarization, and classification~\cite{lu2021codexglue,zhang2024surveylargelanguagemodels}. However, CoderEval-RE and SWEbench-RE cover only the generation category, limiting evaluation to a single modality. To address this limitation, CodeRQ-Bench builds on these two datasets and further introduces Modified-ClassEval-\textbf{R}easoning \textbf{E}valuation (\sumdataset{}) for summarization and Bug Detection-\textbf{R}easoning \textbf{E}valuation (\deftectdataset{}) for classification, each annotated by three experts.

\sumdataset{} is constructed based on Modified-ClassEval~\cite{makharev2025modifiedcodeeval}, a dataset obtained by reformulating ClassEval~\cite{du2023classeval} for code summarization. 
\deftectdataset{} is constructed based on DebugBench~\cite{tian2024debugbench}, a code debugging benchmark in which models determine whether a code snippet contains a defect.
For both datasets, we sample 139 and 252 instances for \sumdataset{} and \deftectdataset{}, respectively, at a confidence level of 90\% and a margin of error of 5\%.
For each sampled instance, we generate a reasoning trace with GPT-4o~\cite{openai2024gpt4o}, selected for its strong overall capability and favorable cost~\cite{openai2025pricing}.
We then label its correctness with three expert annotators.
More details about the construction of \sumdataset{} and \deftectdataset{}, including the used source datasets and sampling procedure, see Appx.~\ref{appx:code_summarization} and ~\ref{appx:defect_detection}.
CodeRQ-Bench's data distribution is shown in Table~\ref{tab:dataset_distribution}.
\begin{table}[ht]
\centering
\caption{CodeRQ-Bench Dataset Distribution}
\vspace{-0.1in}
\footnotesize
\resizebox{=0.9\columnwidth}{!}{%
\renewcommand{\arraystretch}{1.2} 
\begin{tabular}{@{} l l c c @{}} 
\toprule
\textbf{Output Modality} & \textbf{Dataset} & \textbf{Size} & \textbf{\makecell{\# Correct / Incorrect \\ Reasoning}} \\ 
\midrule
\multirow{2}{*}{Code Generation} & CoderEval-RE & 230 & 70 / 160 \\
\cmidrule(lr){2-4}
 & SWEBench-RE & 111 & 21 / 90 \\
\midrule
Code Summarization & ClassEval-RE & 139 & 112 / 27 \\
\midrule
Bug Detection & DebugBench-RE & 252 &  153 / 99 \\
\bottomrule
\end{tabular}%
}
\label{tab:dataset_distribution}
\vspace{-0.1in}
\end{table}
\vspace{-0.1in}

\subsection{Data Annotation Process}
\label{sec:annotation_process}

To ensure annotation consistency, three expert annotators follow a consensus-based workflow, consistent with common dataset annotation practices~\cite{pustejovsky2012natural,Artstein2017iaa,teruel2018increasing,oortwijn2021interrater}. The annotators first jointly establish guidelines for judging reasoning correctness and then independently annotate each instance. To assess reliability before adjudication, we measure inter-annotator agreement during the initial annotation phase using Fleiss' $\kappa$~\cite{fleiss1971measuring}, which is suitable for three annotators. For the two newly constructed datasets, Fleiss' $\kappa$ is 0.91 for \sumdataset{} and 0.95 for \deftectdataset{}, with adjudication rates of 4.32\% and 3.57\%, respectively, indicating almost perfect agreement~\cite{landis1977measurement}.

After independent annotation, annotators review cases with inconsistent labels and assign a consensus label through discussion. If disagreements reveal ambiguities in the guidelines, the rules are refined and the affected instances are re-annotated. This cycle of independent annotation, disagreement resolution, and guideline refinement continues until consensus labels are obtained for all instances.

The annotation guideline is grounded in theories of human problem solving and logical reasoning. Following Pólya’s framework of understanding, planning, execution, and verification~\cite{schoenfeld1987polya}, we operationalize reasoning quality into three stage-specific dimensions: \emph{Comprehension}, \emph{Analysis}, and \emph{Conclusion}. In addition, inspired by Johnson-Laird et al.~\cite{johnson2004reasoning}, which shows that inconsistency detection relies on constructing a coherent mental model, we introduce \emph{Consistency} as a cross-cutting dimension. For each dimension, we define concrete annotation rules for each dataset. The full guidelines for \sumdataset{} and \deftectdataset{} are provided in Appx.~\ref{appx:code_summarization:rules} and Appx.~\ref{appx:defect_detection:rules}.

\section{Analyzing the Limitations of Existing Reasoning Evaluators on Coding Tasks}
\label{sec:analyzing_limitations}
Existing reasoning quality evaluators, including RECEVAL, SOCREVAL, and CaSE~\cite{prasad2023receval,he2024socreval,do2025defines}, perform poorly on coding tasks, often failing to distinguish correct from incorrect reasoning and achieving near-random performance (see Table~\ref{tab:main_result}). To diagnose these limitations and inform the design of improved evaluators, we analyze cases where evaluator scores misalign with the actual reasoning correctness.

\subsection{Analysis Setup and Protocol}
\label{sec:analysis_setup_and_protocol}

\begin{table*}[!h]
\centering
\caption{Taxonomy of existing reasoning-quality evaluator limitations on coding tasks. For each limitation, we report its definition, the distribution of annotated mismatch cases across evaluators, and the breakdown by error type. \textit{REC.}, \textit{SOCR.}, and \textit{CaSE} denote ReCEval, SocREval, and CaSE, respectively; \textit{Missed} and \textit{FA} denote missed errors and false alarms. 
}
\vspace{-0.1in}
\label{tab:limitation_taxonomy}
\scriptsize 
\begin{tabularx}{\textwidth}{@{} 
    >{\raggedright\arraybackslash}p{2cm} 
    >{\raggedright\arraybackslash}X 
    >{\centering\arraybackslash}p{2.5cm} 
    >{\centering\arraybackslash}p{2cm}
    >{\centering\arraybackslash}p{1.7cm} @{}}
\toprule
\textbf{Limitation} & \textbf{Definition} & \textbf{\makecell{Evaluator \\ (\textit{REC.} / \textit{SOCR.} / \textit{CaSE})}}  & \textbf{\makecell{Error Type \\ (\textit{Missed} / \textit{FA})}} & \textbf{\makecell{Representative \\ Case}} \\
\midrule
\midrule

Evidence-Ungrounded Assessment
& The evaluator does not ground its assessment in the evidence needed to verify reasoning correctness in coding tasks, where many claims must be searched and checked against repository context, API documents, or task specifications. 
& 14.4\% / 41.6\% / 44.0\%
& 85.4\% / 14.6\%
& Appx.~\ref{app:case:evidence_ungrounded}
\\
\midrule

Ambiguity Assessment Deficit
& The evaluator cannot determine whether a task is ambiguous or underspecified, nor assess whether the reasoning handles such ambiguity.
& 30.2\% / 36.5\% / 33.3\%
& 77.4\% / 22.6\%
& Appx.~\ref{app:case:ambiguity_deficit}
\\
\midrule

Score Aggregation Failure
&  The evaluator produces an unreliable final score because its scoring rule combines step-level judgments in a way that either lets one misjudged step override an otherwise correct trace or allows fatal correctness errors to be diluted by other steps.
& 70.5\% / -- / 29.5\%
& 24.7\% / 75.3\%
& Appx.~\ref{app:case:score_aggregation}
\\
\midrule

Self-Generated Reference Bias
& The evaluator judges reasoning against self-generated references or standards, creating circular validation in which shared errors or biases are mistaken for correctness. 
& --  / 100\% /  --
& 97.3\% / 2.7\%
& Appx.~\ref{app:case:self_generated_reference}
\\
\midrule

Code-Unaware Assessment
& The evaluator fails to reliably assess code-related reasoning because it cannot correctly represent or judge code-relevant semantics in the reasoning process.
& 100\% / -- / --
& 9.0\% / 91.0\% 
& Appx.~\ref{app:case:code_unaware}
\\
\bottomrule
\end{tabularx}
\vspace{-0.1in}
\end{table*}

We apply existing reasoning evaluators to CodeRQ-Bench and analyze cases where their scores misalign with the ground-truth reasoning correctness. We categorize mismatches into two types: \textit{missed errors}, where incorrect reasoning receives a high score, and \textit{false alarms}, where correct reasoning receives a low score. Since the existing evaluators produce continuous scores, we use the midpoint of each evaluator’s score range as a threshold, treating scores above or equal to the midpoint as high and those below as low. Using this criterion, we identify 1,069 mismatch cases, including 709 \textit{missed errors} and 360 \textit{false alarms}.

We analyze these mismatch cases using a three-step protocol.
First, for each case, we use GPT-5.2~\cite{openai2025gpt52}, which supports long-context, to generate a diagnostic annotation explaining why the evaluator’s score diverges from the ground-truth reasoning correctness. The annotation considers the evaluator’s method description, the task prompt, the LLM-generated reasoning chain and final output, and the evaluator’s detailed assessment outputs.
Second, GPT-5.2 extracts open-coded limitation labels from these diagnostics. Third, we consolidate these labels into higher-level categories through a two-stage process: embedding-based pre-clustering using \texttt{text-embedding-3-large}~\cite{openai2024embedding}, followed by taxonomy consolidation with GPT-5.2, yielding a taxonomy of evaluator limitations.
All prompts are provided in Appx.~\ref{app:implementation_details}. To validate the LLM-based analysis, we randomly sample 65 cases (90\% confidence level, 10\% margin of error) for manual review; 61 cases (93.8\%) contain valid diagnoses and limitation labels.

\subsection{Taxonomy of Limitations in Existing Reasoning Evaluators and Design Insights}
\label{sec:taxonomy_limitations}

\begin{figure}[t!]
\centering
\includegraphics[width=\linewidth]{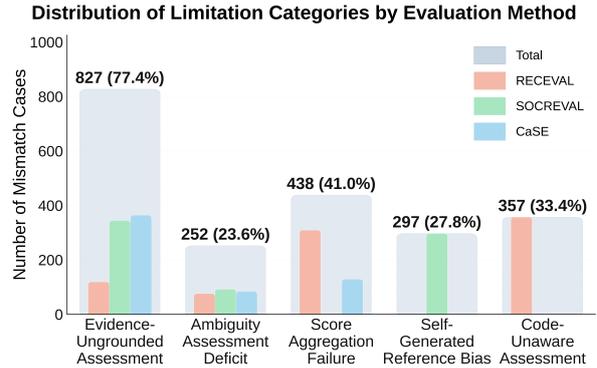}
\caption{Distribution of five limitation categories across 1,069 mismatch cases from three evaluation methods. Background bars show aggregated totals (percentages relative to all cases; categories are non-exclusive). Foreground bars decompose each category by method.}
\label{fig:limitations_dis}
\vspace{-0.1in}
\end{figure}

Based on the mismatch-case analysis above, we organize the observed evaluator limitations into a taxonomy of five limitation categories that cover all mismatch cases.
These categories are non-exclusive, and a single mismatch case may exhibit more than one limitation.
Table~\ref{tab:limitation_taxonomy} summarizes their definitions, along with how each limitation is distributed across evaluators and mismatch types. Fig.~\ref{fig:limitations_dis} visualizes the distribution of limitation annotations across the 1,069 mismatch cases, including a breakdown by evaluation method.
Appx.~\ref{app:limitation_case_studies} presents one representative mismatch case for each category, including the ground-truth diagnosis, the behavior of existing evaluators, and the corresponding assessment by \methodname{}.

\noindent
\textbf{Evidence-Ungrounded Assessment} arises when evaluators score reasoning without verifying claims or assumptions that require external evidence (e.g., repository context, API documentation, or task specifications).
Existing evaluators do not support such evidence retrieval and verification, and therefore tend to accept plausible but unverified reasoning as correct, which results primarily in missed errors (85.4\% of the cases under this limitation).
\begin{itemize}[leftmargin=*]
    \item \textit{Insight 1: Evidence-grounded verification.} 
    CodeRQ evaluators should retrieve and incorporate supporting evidence during assessment (e.g., repository context, API documentation, tests/specifications) and verify whether the reasoning is consistent with that evidence, rather than scoring based on plausibility alone.
\end{itemize}

\noindent
\textbf{Ambiguity Assessment Deficit} arises when the underlying coding problem specification is ambiguous or underspecified, in which case correct reasoning should remain conditional or explicitly uncertain rather than overly certain.
Existing evaluators cannot recognize such ambiguity or assess whether the reasoning responds appropriately to it, and thus often accept overconfident reasoning, resulting primarily in missed errors (77.4\%).
\begin{itemize}[leftmargin=*]
    \item \textit{Insight 2: Ambiguity-aware assessment under underspecification.}
    CodeRQ evaluators should detect ambiguity or underspecification in the task specification and penalize overconfident reasoning when correctness cannot be established.
\end{itemize}

\noindent
\textbf{Score Aggregation Failure} occurs in RECEVAL and CaSE, where the aggregation of step-level scores distorts the final judgment. In RECEVAL, min aggregation makes the score overly sensitive to a single misjudged step, whereas in CaSE, averaging can dilute correctness errors across steps.


\noindent
\textbf{Self-Generated Reference Bias} occurs in SocREVAL, where the evaluator judges reasoning against self-generated references, creating a form of circular validation in which shared errors or biases may be mistaken for correctness. As a result, agreement with these references can reflect shared mistakes rather than true correctness, leading primarily to missed errors (97.3\%).

\begin{itemize}[leftmargin=*]
    \item \textit{Insight 3: Avoid self-referential standards.} 
    CodeRQ evaluators should avoid using self-generated references as the primary standard for correctness; instead, they should prefer evidence-grounded checks or independently sourced verification signals to reduce circular validation and shared-error amplification.
\end{itemize}

\noindent
\textbf{Code-Unaware Assessment} occurs in RECEVAL, whose pipeline converts reasoning steps into premise–conclusion pairs via semantic role labeling (SRL)~\cite{shi2019simple} 
and evaluates entailment using a natural language inference (NLI) model~\cite{laurer2024less}.
Because these models are trained on general natural-language corpora and are brittle on code-mixed reasoning, RECEVAL often misinterprets code-related statements or entailment relations. As a result, many correct reasoning traces are incorrectly flagged, leading predominantly to false alarms (91.0\%).

\begin{itemize}[leftmargin=*]
    \item \textit{Insight 4: Code-aware semantic assessment.} 
    CodeRQ evaluators should incorporate code-aware understanding when assessing reasoning, so that statements about code are evaluated based on the actual semantics and behavior of the code involved.
\end{itemize}

\begin{figure*}[!ht]
\centering
\includegraphics[width=0.9\linewidth]{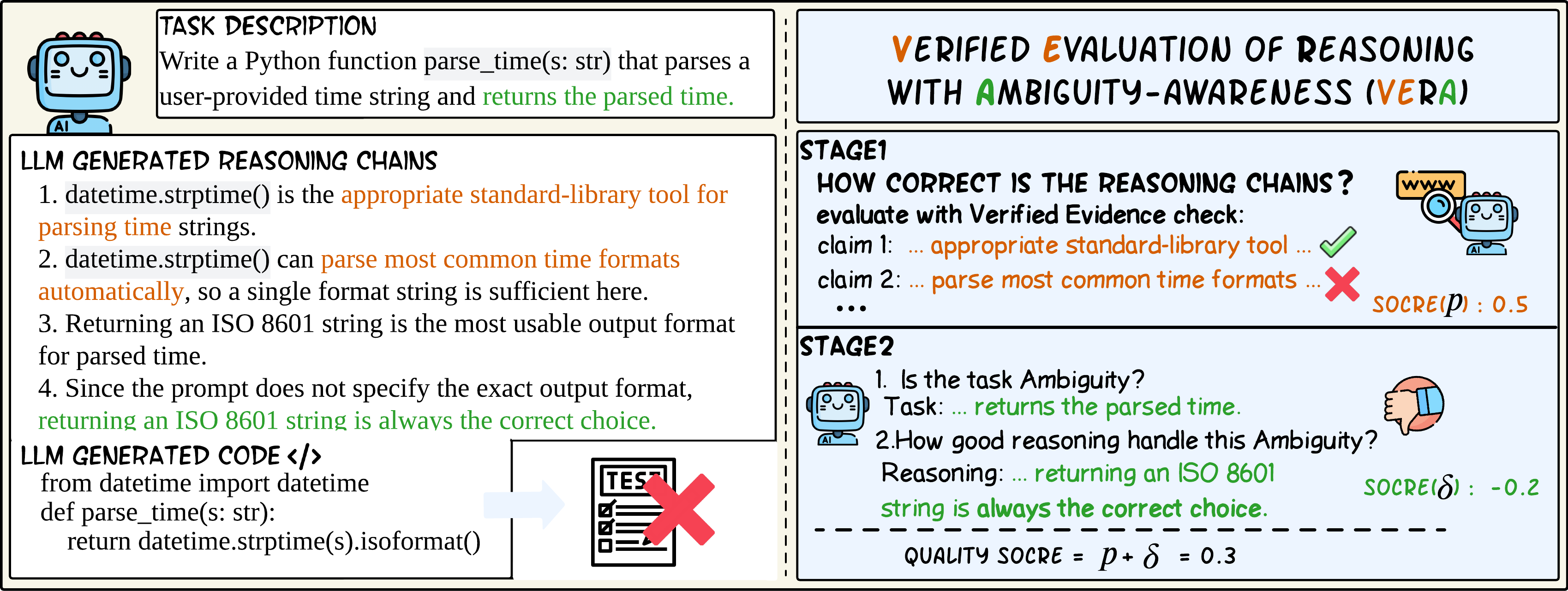}
\caption{Overview of \methodname{}. The left panel shows an input sample with the task description, the LLM-generated reasoning chain, and the generated code output; the right panel shows the two-stage evaluation process. \textcolor{method_orange}{Orange text} marks Stage~1 evidence-verified claims, producing a base score $p$, and \textcolor{method_green}{green text} marks Stage~2 ambiguity-related content, producing a correction term $\delta$. The final score combines $p$ and $\delta$.}

\label{fig:vera_pipeline}
\vspace{-0.1in}
\end{figure*}

\vspace{-0.2in}
\section{Methods}
\label{sec:method}
\vspace{-0.1in}
\subsection{Verified Evaluation of Reasoning with Ambiguity-awareness (\methodname{})}

Guided by the four design insights derived in \S\ref{sec:taxonomy_limitations}, we propose \methodname{}, a two-stage automated evaluation method for assessing the quality of reasoning in coding tasks. Given a task description $q$, a reasoning chain $C=(s_1, s_2, \dots, s_n)$, and a generated output $o$, \methodname{} produces a quality score in $[0,1]$ through two stages (Fig.~\ref{fig:vera_pipeline}).

The first stage performs \emph{verified reasoning evaluation}: an LLM judge scores the correctness of the reasoning trace, with access to a web search tool that it invokes autonomously when encountering technical claims requiring factual verification, and yields a base score $p \in [0,1]$. 
The second stage performs \emph{ambiguity-aware score correction}: a second LLM judge call assesses the degree of task ambiguity and how appropriately the reasoning addresses it, producing a penalty $\delta \le 0$. 
The final score is
$
  \mathrm{\methodname{}}(q,C,o) = \max(p + \delta,\; 0)
$,
where the $\max$ operator ensures that the final score remains in $[0,1]$.

\subsection{Verified Reasoning Evaluation}
\label{sec:stage1}

Motivated by prior LLM-as-a-Judge studies~\cite{zheng2023judging,gu2024survey}, we use an LLM judge as a starting point for automated reasoning evaluation in this setting.
Given a sample $(q, C, o)$, the first stage prompts an LLM judge to assess the overall correctness of the reasoning chain $C$, focusing on whether the reasoning captures the technical requirements of the task, makes technically sound claims, and supports an output that correctly addresses the problem, in line with Insight~4 (\S\ref{sec:taxonomy_limitations}). 
The full evaluation prompt and rubric are provided in Appx.~\ref{app:verified_prompt}. 

To reduce reliance on the judge's parametric knowledge, the judge is given access to a search tool that invokes autonomously during evaluation, in line with Insights~1 and~3 (\S\ref{sec:taxonomy_limitations}). When the reasoning chain contains verifiable technical claims, such as the behavior of a standard library function or the documented behavior of a framework component, the judge issues a search query to retrieve external evidence before rendering its judgment. The retrieved evidence is then used in the subsequent evaluation, allowing the judge to compare the claim against external sources before producing its final score. This mechanism shifts evaluation from pure plausibility-based assessment toward evidence-grounded verification; its empirical effect is measured via an ablation study in \S\ref{sec:ablation}.

The judge produces an integer score $r \in [1, 10]$ according to a rubric with anchored descriptions ranging from fundamentally incorrect reasoning ($r{=}1$) to fully correct reasoning, with verifiable claims correctly supported when applicable ($r{=}10$). The base score is obtained by normalizing to the unit interval:
$
  p = \frac{r - 1}{9}
  \label{eq:normalize}
$.

\subsection{Ambiguity-Aware Score Correction}
\label{sec:stage2}

The base score $p$ from Stage~1 evaluates whether the reasoning is correct and, when needed, grounds this judgment in verifiable evidence. However, like existing reasoning evaluators, it does not assess whether the task itself is ambiguous, nor whether the reasoning appropriately responds to such ambiguity. Stage~2 addresses this limitation by evaluating both the degree of task ambiguity and the quality of ambiguity handling, and applies a penalty when ambiguity is handled poorly, in line with Insight~2 (\S\ref{sec:taxonomy_limitations}).

In Stage~2, we also use an LLM judge to produce two judgments for each sample $(q, C, o)$: an \emph{ambiguity level} $a \in [0,1]$, measuring how underspecified the task $q$ is, and a \emph{handling quality} score $h \in [0,1]$, measuring how appropriately the reasoning chain $C$ handles that ambiguity. Both are elicited in a single call to ensure that they are assessed under the same interpretation of the instance. The full prompt and calibrated rubrics are provided in Appx.~\ref{app:ambiguity_prompt}.

The two scores are combined into a penalty $\delta$:
\[
\delta =
\begin{cases}
a \cdot \min(2h-1,0), & \text{if } a \ge \tau,\\
0, & \text{otherwise,}
\end{cases}
\]
where $\tau$ is a threshold on the ambiguity level.
The transformation $2h{-}1$ recenters $h$ so that $h{=}0.5$ serves as the neutral point, while $h{<}0.5$ maps to a negative value.
The $\min(\cdot, 0)$ operator ensures that $\delta \le 0$: good ambiguity handling prevents penalty but does not inflate the score beyond what the correctness assessment already warrants.
Scaling by $a$ makes the penalty magnitude proportional to the degree of ambiguity.
$\tau$ is set to $0.4$; a sensitivity analysis is provided in Appx.~\ref{app:tau_ablation}.

\begin{table*}[!ht]
\caption{Performance of \methodname{} and six baselines on the four CodeRQ-Bench datasets, evaluated using Somers'~D, Spearman's~$\rho$, AUCROC, and AUPRC ($\uparrow$ higher is better). Best results highlighted in bold.}
\vspace{-0.1in}
\centering
\setlength{\tabcolsep}{3pt}
\resizebox{\textwidth}{!}{%
\begin{tabular}{lcccc|cccc}
\toprule
\toprule
\multicolumn{1}{c}{\multirow{2}{*}{\textbf{Method}}} & \multicolumn{4}{c}{\textbf{CoderEval-RE} (Generation)} & \multicolumn{4}{c}{\textbf{SWEbench-RE} (Generation)} \\ \cmidrule{2-9}
\multicolumn{1}{c}{} & Somers' D $\uparrow$ & Spearman's $\rho$ $\uparrow$ & AUCROC $\uparrow$ & AUPRC $\uparrow$ & Somers' D $\uparrow$ & Spearman's $\rho$ $\uparrow$ & AUCROC $\uparrow$ & AUPRC $\uparrow$ \\ \midrule

\rowcolor{tablegray}
RECEVAl      & 0.1255 & 0.1000 (p=0.1308) & 0.5627 & 0.3516 & 0.0878 & 0.0596 (p=0.5344) & 0.5439 & 0.2125 \\ 
SOCREVAl     & 0.1399 & 0.1248 (p=0.0588) & 0.5700 & 0.3410 & 0.1556 & 0.1830 (p=0.0545) & 0.5778 & 0.2165 \\ 
\rowcolor{tablegray}
CaSE         & 0.0368 & 0.0392 (p=0.5543) & 0.5184 & 0.3131 & -0.0243& -0.0588 (p=0.5400) & 0.4878 & 0.1855 \\ 
MAD$^{*}$          & -0.0408 & -0.0480 (p=0.4684) & 0.4796 & 0.2963 & -0.0746 & -0.0644 (p=0.5019) & 0.4627 & 0.1785 \\
\rowcolor{tablegray}
LLM-as-Judge & 0.0864 & 0.1077 (p=0.1032) & 0.5432 & 0.3252 & 0.0222 & 0.0280 (p=0.7707) & 0.5111 & 0.1924 \\ 
AutoRace$^{*}$ & 0.0178 & 0.0146 (p=0.8253) & 0.5089 & 0.3359 & 0.0344 & 0.0247 (p=0.7972) & 0.5172 & 0.2114 \\
\rowcolor{lightbrown}
\textbf{\methodname{}} (Ours) & \textbf{0.3811} & \textbf{0.3076} (p<0.0001) & \textbf{0.6905} & \textbf{0.4615} & \textbf{0.2799} & \textbf{0.1947} (p=0.0406) & \textbf{0.6399} & \textbf{0.3058} \\ 

\midrule
\midrule
\multicolumn{1}{c}{\multirow{2}{*}{\textbf{Method}}} & \multicolumn{4}{c}{\textbf{\sumdataset{}} (Summarization)} & \multicolumn{4}{c}{\textbf{\deftectdataset{}} (Classification)} \\ \cmidrule{2-9}
\multicolumn{1}{c}{} & Somers' D $\uparrow$ & Spearman's $\rho$ $\uparrow$ & AUCROC $\uparrow$ & AUPRC $\uparrow$ & Somers' D $\uparrow$ & Spearman's $\rho$ $\uparrow$ & AUCROC $\uparrow$ & AUPRC $\uparrow$ \\ \midrule
\rowcolor{tablegray}
RECEVAl      & -0.0218& -0.0150 (p=0.8613)& 0.4891 & 0.8239 & -0.0896& -0.0758 (p=0.2306)& 0.4552 & 0.5844 \\ 
SOCREVAl     & 0.1680 & 0.1814 (p=0.0330) & 0.5840 & 0.8345 & 0.1819 & 0.1958 (p=0.0018) & 0.5909 & 0.6566 \\ 
\rowcolor{tablegray}
CaSE         & 0.0397 & 0.0582 (p=0.4960) & 0.5198 & 0.8120 & 0.3538 & 0.3402 (p<0.0001) & 0.6769 & 0.7431 \\ 
MAD$^{*}$          & 0.2209 & 0.1588 (p=0.0619) & 0.6104 & 0.8494 & -0.0166 & -0.0197 (p=0.7557) & 0.4917 & 0.6192 \\
\rowcolor{tablegray}
LLM-as-Judge & -0.0089 & -0.0418 (p=0.6252) & 0.4955 & 0.8044 & 0.0530 & 0.0471 (p=0.4569) & 0.5265 & 0.6262 \\ 
AutoRace$^{*}$ & 0.2500 & 0.1905 (p=0.0247) & 0.6250 & 0.8502 & 0.2063 & 0.1772 (p=0.0048) & 0.6032 & 0.7142 \\
\rowcolor{lightbrown}
\textbf{\methodname{}} (Ours) & \textbf{0.4180} & \textbf{0.3085} (p=0.0002) & \textbf{0.7090} & \textbf{0.8869} & \textbf{0.4351} & \textbf{0.3845} (p<0.0001) & \textbf{0.7176} & \textbf{0.7939} \\ 

\bottomrule
\bottomrule
\end{tabular}%
}
\label{tab:main_result}
\vspace{-0.1in}
\end{table*}

\section{Experiments}
\subsection{Experimental Setup}

\paragraph{Methods Compared.}
We compare \methodname{} against six representative reasoning evaluation baselines. 
\textbf{RECEVAL}~\cite{prasad2023receval} is a step-level evaluator that measures intra-step correctness, inter-step correctness, and informativeness, and aggregates them into chain-level scores. 
\textbf{SOCREVAL}~\cite{he2024socreval} is a Socratic-prompting-based LLM judge that assigns a discrete overall score to the full reasoning trace. 
\textbf{CaSE}~\cite{do2025defines} is a causal stepwise evaluator that scores each reasoning step in terms of relevance and coherence using only its preceding context. 
\textbf{LLM-as-Judge}~\cite{zheng2023judging} is a vanilla LLM-based evaluator that directly assigns an overall quality score to the reasoning trace. 
\textbf{MAD$^{*}$}~\cite{zhang2025they} is adapted from MAD, a debate-based framework for detecting low-quality reasoning, with its original binary output converted into a continuous score in $[0,1]$ for comparison.
\textbf{AutoRace$^{*}$}~\cite{hao2024new} is adapted from AutoRace, a task-adaptive automated reasoning evaluation framework that induces task-specific criteria from example incorrect reasoning chains and then uses an LLM judge for assessment, with its original binary output converted into a continuous score in $[0,1]$ for comparison.
Since MAD$^{*}$ relies on role-based debate and AutoRace$^{*}$ depends on induced task-specific criteria rather than fixed evaluator formulations, we do not include them in the limitation analysis in \S\ref{sec:taxonomy_limitations}. 
Additional details of the baseline methods are provided in Appx.~\ref{app:baseline_overview}.

\paragraph{Evaluation Metrics.}
We evaluate each reasoning evaluator using rank-based and discriminative metrics: Somers'~D~\cite{somers1962new}, Spearman's~$\rho$~\cite{spearman1961proof}, AUCROC~\cite{green1966signal}, and AUPRC~\cite{davis2006relationship}. 
Together, these metrics assess both ranking quality and discriminative ability. Full details are provided in Appx.~\ref{app:metrics}.

\paragraph{Implementation Details.}
RECEVAL is implemented following its original configuration with three dimensions averaged into a unified score. 
SOCREVAL and CaSE use GPT-4 by default and GPT-4.1 on SWEbench-RE due to context length limits.
VERA, MAD$^{*}$, LLM-as-Judge, and AutoRace$^{*}$ use GPT-4o-mini for cost efficiency.
VERA uses $\tau=0.4$ based on the sensitivity study in Appx.~\ref{app:tau_ablation}, and all outputs are normalized to $[0,1]$. To ensure the robustness of our findings, we repeat each experiment three times and report the average performance. 
Full details are in Appx.~\ref{app:implementation}.

\subsection{Results}

\paragraph{Cross-dataset comparison and the value of CodeRQ-Bench.}
Table~\ref{tab:main_result} reports Somers'~D and Spearman's~$\rho$ measuring the association between predicted scores and ground-truth correctness labels, respectively. AUCROC measures the probability that a correct reasoning trace receives a higher score than an incorrect one, while AUPRC is particularly informative under class imbalance. 

The results show that baseline performance varies across coding settings. SOCREVAL is the strongest baseline on the two generation-oriented datasets, but this advantage does not generalize to summarization and classification tasks.
CaSE becomes more competitive on bug detection, whereas MAD$^{*}$ and AutoRace$^{*}$ perform relatively better on code summarization than on generation. These shifts indicate that conclusions drawn from generation-only benchmarks provide an incomplete view of evaluator performance. 

Notably, even \methodname{} does not achieve uniform performance across settings, underscoring the need for evaluation beyond generation-only benchmarks. Overall, these results highlight the importance of CodeRQ-Bench’s broad task coverage for assessing reasoning evaluators in coding tasks.

\paragraph{Overall performance of \methodname{}.}
Table~\ref{tab:main_result} shows that \methodname{} consistently achieves the best performance across all four datasets and all evaluation metrics, indicating substantially stronger alignment with reasoning correctness in both rank consistency and discriminative ability. The gains are particularly pronounced on the generation-oriented datasets. On CoderEval-RE, Somers'~D increases from 0.1399 to 0.3811 and AUCROC from 0.5700 to 0.6905; on SWEbench-RE, Somers'~D improves from 0.1556 to 0.2799 and AUPRC from 0.2165 to 0.3058. These results suggest that existing evaluators are particularly limited when reasoning makes factual claims about code behavior, API usage, or library semantics that cannot be verified from the reasoning text alone.

A similar advantage holds on \sumdataset{} (0.4180 Somers'~D, 0.7090 AUCROC) and \deftectdataset{} (0.4351 Somers'~D, 0.7176 AUCROC), where CaSE is the strongest baseline yet still falls below \methodname{} on all metrics, indicating that step-level relevance and coherence alone are insufficient for accurate reasoning evaluation.

LLM-as-Judge yields modest and sometimes unstable performance, including negative rank correlations on some datasets.
MAD$^{*}$ is relatively competitive on \sumdataset{} (0.2209 Somers'~D, 0.6104 AUCROC) but performs poorly on the generation and bug detection datasets. Together, these results show that strong reasoning evaluation does not emerge from a general-purpose LLM judge or debate-based procedure alone; it requires a design that explicitly accounts for evidence grounding and task ambiguity. Notably, despite using the more economical \texttt{GPT-4o-mini}, \methodname{} outperforms GPT-4-based baselines by a large margin, confirming that the gains stem from the evaluation design rather than model size.

\paragraph{Correlation between Reasoning-Evaluator Scores and Final-Answer Correctness.}
\label{sec:final_output_correctness}

Table~\ref{tab:final_output_correctness} reports how the scores produced by \methodname{} and the baselines correlate with the actual correctness of the final generated answer.
\methodname{} achieves the strongest score-correctness association on both datasets and ranks first across all four metrics. Its Somers'~D and AUCROC are 0.3811 and 0.6905 on CoderEval-RE, and 0.2819 and 0.6410 on \deftectdataset{}, respectively.
These results indicate that \methodname{}'s reasoning-evaluator scores align more closely with final-answer correctness than the baseline scores.

\begin{table*}[!t]
\caption{Association between reasoning-evaluator scores and final-output correctness on CoderEval-RE and DebugBench-RE, evaluated using Somers'~D, Spearman's~$\rho$, AUCROC, and AUPRC ($\uparrow$ higher is better). Best results are highlighted in bold.}
\vspace{-0.1in}
\centering
\setlength{\tabcolsep}{3pt}
\resizebox{\textwidth}{!}{%
\begin{tabular}{lcccc|cccc}
\toprule
\toprule
\multicolumn{1}{c}{\multirow{2}{*}{\textbf{Method}}} & \multicolumn{4}{c}{\textbf{CoderEval-RE} } & \multicolumn{4}{c}{\textbf{DebugBench-RE} } \\ \cmidrule{2-9}
\multicolumn{1}{c}{} & Somers' D $\uparrow$ & Spearman's $\rho$ $\uparrow$ & AUCROC $\uparrow$ & AUPRC $\uparrow$ & Somers' D $\uparrow$ & Spearman's $\rho$ $\uparrow$ & AUCROC $\uparrow$ & AUPRC $\uparrow$ \\ \midrule

\rowcolor{tablegray}
RECEVAL      & 0.1255 & 0.1000 (p=0.1308) & 0.5627 & 0.3516 & -0.0843 & -0.0539 (p=0.3945) & 0.4579 & 0.8176 \\
SOCREVAL     & 0.1399 & 0.1248 (p=0.0588) & 0.5700 & 0.3410 & 0.1233 & 0.1004 (p=0.1120) & 0.5617 & 0.8566 \\
\rowcolor{tablegray}
CaSE         & 0.0368 & 0.0392 (p=0.5543) & 0.5184 & 0.3131 & 0.0822 & 0.0597 (p=0.3450) & 0.5411 & 0.8628 \\
MAD$^{*}$    & -0.0408 & -0.0480 (p=0.4684) & 0.4796 & 0.2963 & -0.0342 & -0.0307 (p=0.6273) & 0.4829 & 0.8393 \\
\rowcolor{tablegray}
LLM-as-Judge & 0.0864 & 0.1077 (p=0.1032) & 0.5432 & 0.3252 & 0.0504 & 0.0338 (p=0.5931) & 0.5252 & 0.8441 \\
AutoRace$^{*}$ & 0.0178 & 0.0146 (p=0.8253) & 0.5089 & 0.3359 & 0.1835 & 0.1194 (p=0.0590) & 0.5918 & 0.8771 \\
\rowcolor{lightbrown}
\textbf{\methodname{}} (Ours) & \textbf{0.3811} & \textbf{0.3076} (p<0.0001) & \textbf{0.6905} & \textbf{0.4615} & \textbf{0.2819} & \textbf{0.1883} (p=0.0027) & \textbf{0.6410} & \textbf{0.9007} \\

\bottomrule
\bottomrule
\end{tabular}%
}
\label{tab:final_output_correctness}
\vspace{-0.1in}
\end{table*}

\subsubsection{Ablation Study}
\label{sec:ablation}
\begin{table}[!t]
\caption{Ablation study of \methodname{} on four CodeRQ-Bench datasets, including two ablated variants: \textit{w/o} verif. (without evidence verification) and \textit{w/o} ambig. (without ambiguity-aware correction).}
\vspace{-0.1in}
\centering
\setlength{\tabcolsep}{3pt}
\resizebox{\columnwidth}{!}{%
\begin{tabular}{cl|cccc}
\toprule
\toprule
\textbf{Dataset} & \multicolumn{1}{c}{\textbf{Method}} & Somers' D $\uparrow$ & Spearman's $\rho$ $\uparrow$ & AUCROC $\uparrow$ & AUPRC $\uparrow$ \\ \midrule
\multirow{3}{*}{\begin{tabular}[c]{@{}c@{}}\textbf{CoderEval-RE}\\
(Generation) \end{tabular}}
 & \cellcolor{lightbrown}\textbf{\methodname{}} (Ours) & \cellcolor{lightbrown}\textbf{0.3811} & \cellcolor{lightbrown}\textbf{0.3076} (p<0.0001) & \cellcolor{lightbrown}\textbf{0.6905} & \cellcolor{lightbrown}\textbf{0.4615} \\
 & \quad \textit{w/o} verif. & 0.2945 & 0.2369 (p=0.0003) & 0.6472 & 0.4356 \\
 & \cellcolor{tablegray}\quad \textit{w/o} ambig. & \cellcolor{tablegray}0.3216 & \cellcolor{tablegray}0.2658 (p<0.0001) & \cellcolor{tablegray}0.6608 & \cellcolor{tablegray}0.4364 \\ \midrule
\multirow{3}{*}{\begin{tabular}[c]{@{}c@{}}\textbf{SWEbench-RE}\\
(Generation) \end{tabular}}
 & \cellcolor{lightbrown}\textbf{\methodname{}} (Ours) & \cellcolor{lightbrown}\textbf{0.2799} & \cellcolor{lightbrown}\textbf{0.1947} (p=0.0406) & \cellcolor{lightbrown}\textbf{0.6399} & \cellcolor{lightbrown}\textbf{0.3058} \\
 & \quad \textit{w/o} verif. & -0.0376 & -0.0261 (p=0.7855) & 0.4812 & 0.1842 \\
 & \cellcolor{tablegray}\quad \textit{w/o} ambig. & \cellcolor{tablegray}0.2571 & \cellcolor{tablegray}0.1796 (p=0.0593) & \cellcolor{tablegray}0.6286 & \cellcolor{tablegray}0.3012 \\ \midrule
\multirow{3}{*}{\begin{tabular}[c]{@{}c@{}}\textbf{\sumdataset{}}\\
(Summarization) \end{tabular}}
 & \cellcolor{lightbrown}\textbf{\methodname{}} (Ours) & \cellcolor{lightbrown}\textbf{0.4180} & \cellcolor{lightbrown}\textbf{0.3085} (p=0.0002) & \cellcolor{lightbrown}\textbf{0.7090} & \cellcolor{lightbrown}\textbf{0.8869} \\
 & \quad \textit{w/o} verif. & 0.2622 & 0.1984 (p=0.0192) & 0.6311 & 0.8561 \\
 & \cellcolor{tablegray}\quad \textit{w/o} ambig. & \cellcolor{tablegray}0.4134 & \cellcolor{tablegray}0.3051 (p=0.0003) & \cellcolor{tablegray}0.7067 & \cellcolor{tablegray}0.8864 \\ \midrule
\multirow{3}{*}{\begin{tabular}[c]{@{}c@{}}\textbf{\deftectdataset{}}\\ (Classification)\end{tabular}}
 & \cellcolor{lightbrown}\textbf{\methodname{}} (Ours) & \cellcolor{lightbrown}\textbf{0.4351} & \cellcolor{lightbrown}\textbf{0.3845} (p<0.0001) & \cellcolor{lightbrown}\textbf{0.7176} & \cellcolor{lightbrown}\textbf{0.7939} \\
 & \quad \textit{w/o} verif. & 0.3440 & 0.3094 (p<0.0001) & 0.6720 & 0.7575 \\
 & \cellcolor{tablegray}\quad \textit{w/o} ambig. & \cellcolor{tablegray}0.4184 & \cellcolor{tablegray}0.3725 (p<0.0001) & \cellcolor{tablegray}0.7092 & \cellcolor{tablegray}0.7887 \\ \bottomrule
\bottomrule
\end{tabular}%
}
\label{tab:ablation_study}
\vspace{-0.15in}
\end{table}

Table~\ref{tab:ablation_study} reports two ablated variants: \textit{w/o} verif., in which the Stage~1 evaluator judges reasoning correctness without externally verifying verifiable technical claims, and \textit{w/o} ambig., which removes the ambiguity-aware correction.
Both consistently underperform the full model, confirming that each component is necessary.
Removing evidence verification causes the largest degradation, particularly on the generation-oriented datasets: Somers'~D drops from 0.3811 to 0.2945 on CoderEval-RE and from 0.2799 to $-$0.0376 on SWEbench-RE, with similar trends on \sumdataset{} and \deftectdataset{}. 
This identifies evidence verification as the dominant driver of improvement, especially when reasoning makes factual claims about code behavior, API usage, or library semantics that require external grounding to assess.
Removing ambiguity-aware correction yields a smaller but consistent drop, most visible on the generation-oriented datasets where task requirements are more often underspecified. Its impact on \sumdataset{} and \deftectdataset{} is marginal, confirming that this component provides complementary benefits primarily under ambiguous task conditions.

\paragraph{Formula ablation.}
We further test whether the specific ambiguity-aware correction in \S\ref{sec:stage2} is necessary while holding the Stage~1 and Stage~2 outputs fixed. We compare the full formulation with three alternatives: \textit{no correction} sets $\delta{=}0$; \textit{no threshold} applies $\delta{=}a\min(2h{-}1,0)$ to every task; and \textit{no ambiguity scaling} applies $\delta{=}\mathbf{1}[a\geq\tau]\min(2h{-}1,0)$.

\begin{table}[!t]
\caption{Formula ablation averaged across the four CodeRQ-Bench datasets. ``Best baseline'' is the best of the six baselines for each metric.}
\vspace{-0.1in}
\centering
\setlength{\tabcolsep}{3.2pt}
\resizebox{\columnwidth}{!}{%
\begin{tabular}{lcccc}
\toprule
\toprule
\textbf{Method} & Somers' D $\uparrow$ & Spearman's $\rho$ $\uparrow$ & AUCROC $\uparrow$ & AUPRC $\uparrow$ \\
\midrule
Best baseline & 0.1613 & 0.1713 & 0.5807 & 0.5279 \\
\midrule
\cellcolor{lightbrown}\textbf{\methodname{}} (full) & \cellcolor{lightbrown}\textbf{0.3785} & \cellcolor{lightbrown}\textbf{0.2988} & \cellcolor{lightbrown}\textbf{0.6893} & \cellcolor{lightbrown}\textbf{0.6120} \\
\quad No correction & 0.3526 & 0.2807 & 0.6763 & 0.6032 \\
\cellcolor{tablegray}\quad No threshold & \cellcolor{tablegray}0.3712 & \cellcolor{tablegray}0.2928 & \cellcolor{tablegray}0.6856 & \cellcolor{tablegray}0.6084 \\
\quad No ambiguity scaling & 0.3784 & 0.2988 & 0.6892 & 0.6111 \\
\bottomrule
\bottomrule
\end{tabular}%
}
\label{tab:formula_ablation}
\vspace{-0.15in}
\end{table}

Table~\ref{tab:formula_ablation} shows that the full formulation performs best or ties for best on all four metrics averaged across CodeRQ-Bench. Removing the correction causes the largest decline, confirming the value of ambiguity-aware adjustment. Removing the threshold also degrades every metric, while removing ambiguity scaling produces nearly identical rank correlations but slightly lower discriminative performance. Even the weakest variant remains above the strongest baseline on every metric, indicating that the gains are not tied to a single heuristic choice.

\section{Conclusion}
We introduce CodeRQ-Bench and analyze 1,069 mismatches from existing reasoning-quality evaluators, deriving four design insights. Next, we propose VERA, a two-stage evaluator combining evidence-grounded verification with ambiguity-aware score correction. Our results show that VERA outperforms baselines across four datasets, highlighting the need for both broader benchmarks and evaluation methods beyond plausibility-based judgment to reliably assess reasoning in coding.




\section*{Limitations}
\vspace{-0.1in}
Despite its contributions, this work has certain limitations. Although CodeRQ-Bench provides broad coverage across generation, summarization, and classification tasks, the current benchmark still focuses on relatively controlled evaluation settings and does not yet fully capture more realistic coding workflows, such as tool-integrated problem solving, interactive debugging, and richer long-context repository reasoning. Future work can extend the benchmark to these broader settings, enabling more comprehensive evaluation of reasoning quality in coding tasks.

\section*{Ethical considerations}
\vspace{-0.1in}
This work adheres to ethical standards emphasizing transparency, reliability, and privacy in reasoning-quality evaluation for coding tasks. By openly releasing CodeRQ-Bench, reporting our empirical analysis of existing evaluator limitations, and introducing VERA, this work provides a standardized foundation for advancing more reliable assessment of LLM reasoning in coding tasks. All benchmark instances are derived from publicly available datasets and contain no personally identifiable information. External models and APIs are used in accordance with their respective terms of service. Additionally, we used ChatGPT exclusively to improve minor grammar in the final manuscript text.

\section*{Broader Impacts}
\vspace{-0.1in}
The benchmark and evaluator proposed in this paper provide a benchmark foundation for reasoning-quality evaluation in coding tasks. By standardizing the evaluation setting and moving beyond final-output correctness, this work supports progress in reliable coding assistants, reasoning evaluation methods, and trustworthy LLM research for software engineering. The benchmark promotes transparency, reproducibility, and further innovation, ultimately contributing to more reliable and responsible use of LLMs in coding-related applications.

\section*{Acknowledgments}
\vspace{-0.1in}
This work was supported in part by the National Science Foundation (NSF) under Award No. CCF-2326489.



\bibliography{custom}




\appendix

\section*{Supplementary Material}

\setcounter{section}{0}
\setcounter{figure}{0}
\setcounter{table}{0}
\makeatletter 
\renewcommand{\thesection}{\Alph{section}}
\renewcommand{\theHsection}{\Alph{section}}
\renewcommand{\thefigure}{A\arabic{figure}} 
\renewcommand{\theHfigure}{A\arabic{figure}} 
\renewcommand{\thetable}{A\arabic{table}}
\renewcommand{\theHtable}{A\arabic{table}}
\makeatother

\renewcommand{\thetable}{A\arabic{table}}
\setcounter{equation}{0}
\renewcommand{\theequation}{A\arabic{equation}}

\section{CodeRQ-Bench's Dataset Details}
\subsection{CoderEval-RE and SWEbench-RE}
\label{appx:dataset_details}
Zhang et al.~\cite{zhang2025they} evaluate three LLMs on CoderEval~\cite{yu2024codereval} and SWE-bench~\cite{jimenez2024swebench}, providing expert annotations that verify the correctness of LLM-generated reasoning traces.

Due to the partial release of the original experimental data, we use the expert-annotated reasoning traces generated by DeepSeek-R1~\cite{guo2025deepseek} to ensure the completeness and reproducibility of our baseline.

\paragraph{CoderEval-RE}
CoderEval~\cite{yu2024codereval} is a benchmark designed for pragmatic code generation, containing real-world coding tasks extracted from open-source projects. CoderEval-RE extends this benchmark with reasoning trace annotations, enabling evaluation of the logical correctness of LLM reasoning processes during code generation.

\paragraph{SWEbench-RE}
SWE-bench~\cite{jimenez2024swebench} consists of real GitHub issues and their corresponding pull requests from popular Python repositories. SWEbench-RE augments this benchmark with expert-annotated reasoning traces, allowing assessment of whether LLMs reason correctly when resolving software engineering tasks.

\subsection{\sumdataset{} Dataset Details}
\label{appx:code_summarization}

\subsubsection{Source Dataset - Modified-ClassEval}
ClassEval~\cite{du2023classeval} is a class-level Python code generation benchmark consisting of 100 manually constructed class-level programming tasks, covering 100 classes and 410 methods. Modified-ClassEval adapts this benchmark for the code summarization setting by extracting the contextual information, function implementation, and corresponding natural language summary for each method.

\subsubsection{Construction Details}
To construct \sumdataset{}, we first filtered the 410 methods to remove cases containing fewer than five lines of code, as such trivial implementations typically do not require substantial reasoning for summarization. This filtering step resulted in a pool of 282 candidate methods. From this pool, we randomly sampled instances using a confidence level of 90\% and a margin of error of 5\%, yielding a final dataset of 139 methods.
For the sampled instances, we use GPT-4o to generate reasoning traces.
The correctness of these reasoning traces is then annotated by three expert annotators following the annotation process described in \S ~\ref{sec:annotation_process}.

\subsubsection{Annotation Rules}
\label{appx:code_summarization:rules}
Table~\ref{tab:sumdataset_rules} summarizes the dimension-specific annotation rules for \sumdataset{}.
\begin{table}[ht]
\centering
\caption{Annotation rules for \sumdataset{} (Code Summarization) across four evaluation dimensions.}
\footnotesize 
\renewcommand{\arraystretch}{1.3} 
\label{tab:sumdataset_rules}
\begin{tabularx}{\columnwidth}{@{} c X @{}}
\toprule
\multicolumn{2}{@{}l}{\textbf{Dimension: Comprehension}} \\
\midrule
\textbf{R1} & The reasoning must correctly identify the input parameters and return values of the code. \\
\textbf{R2} & The reasoning must accurately trace the control flow (loops, conditionals, branches) of the code. \\
\textbf{R3} & The reasoning must correctly interpret the operations performed on data structures. \\
\textbf{R4} & The reasoning must not attribute functionality that does not exist in the code. \\

\midrule
\multicolumn{2}{@{}l}{\textbf{Dimension: Analysis}} \\
\midrule
\textbf{R1} & The reasoning must correctly map variable names and function calls to their semantic roles. \\
\textbf{R2} & The reasoning must abstract implementation details to appropriate conceptual descriptions. \\
\textbf{R3} & The reasoning must correctly identify the algorithmic or design pattern employed in the code. \\

\midrule
\multicolumn{2}{@{}l}{\textbf{Dimension: Conclusion}} \\
\midrule
\textbf{R1} & The reasoning must address the primary functionality of the code. \\
\textbf{R2} & The reasoning must cover all significant code branches and edge cases that affect the summary. \\
\textbf{R3} & The reasoning must not include irrelevant analysis that does not contribute to understanding the code's purpose. \\

\midrule
\multicolumn{2}{@{}l}{\textbf{Dimension: Consistency}} \\
\midrule
\textbf{R1} & The reasoning must not contain self-contradictory statements about code behavior. \\
\textbf{R2} & Each reasoning step must logically follow from the previous step or from the code itself. \\
\textbf{R3} & The final summary must be logically derivable from the reasoning steps presented. \\
\bottomrule
\end{tabularx}
\end{table}

\subsection{\deftectdataset{} Dataset Details}
\label{appx:defect_detection}

\subsubsection{Source Dataset - DebugBench}
DebugBench~\cite{tian2024debugbench} is a bug detection benchmark composed of LeetCode-style code snippets in which bugs are injected using GPT-4 and subsequently verified by both human annotators and LLMs. The dataset contains both correct and buggy programs spanning four bug categories: syntax errors, reference errors, logic errors, and multiple errors (a combination of the previous three categories).

\subsubsection{Construction Details}
To construct \deftectdataset{}, we filtered the dataset to remove instances labeled solely as syntax errors, as detecting syntax errors is largely trivial and does not require substantive reasoning. After filtering, 3,492 instances remained. From this pool, we randomly sampled examples using a confidence level of 90\% and a margin of error of 5\%, resulting in a final dataset of 252 methods, consisting of 126 bug-free and 126 buggy programs.
For the sampled instances, we use GPT-4o to generate reasoning traces.
The correctness of these reasoning traces is then annotated by three expert annotators following the annotation process described in \S ~\ref{sec:annotation_process}.

\subsubsection{Annotation Rules}
\label{appx:defect_detection:rules}
Table~\ref{tab:deftectdataset_rules} summarizes the dimension-specific annotation rules for \deftectdataset{}.

\begin{table}[ht]
\centering
\caption{Annotation rules for \deftectdataset{} (Bug Detection) across four evaluation dimensions.}
\footnotesize 
\renewcommand{\arraystretch}{1.3} 
\label{tab:deftectdataset_rules}
\begin{tabularx}{\columnwidth}{@{} c X @{}}
\toprule
\multicolumn{2}{@{}l}{\textbf{Dimension: Comprehension}} \\
\midrule
\textbf{R1} & The reasoning must correctly identify parameters, their types (if inferable), and their roles where bugs are identified. \\
\textbf{R2} & The reasoning must correctly identify the return value(s), including affected conditionals where bugs are identified. \\
\textbf{R3} & The reasoning must correctly describe how bug-related data structures (e.g., lists, maps, trees) are read, modified, or traversed. \\

\midrule
\multicolumn{2}{@{}l}{\textbf{Dimension: Analysis}} \\
\midrule
\textbf{R1} & All bugs that affect observable behavior must be acknowledged. \\
\textbf{R2} & The reasoning must correctly describe the execution order of conditionals, loops, and function calls. \\
\textbf{R3} & The reasoning must not describe conditionals that are unreachable or nonexistent in the code. \\
\textbf{R4} & The reasoning must not misattribute algorithmic behavior (e.g., sorting, filtering) where it does not occur. \\

\midrule
\multicolumn{2}{@{}l}{\textbf{Dimension: Conclusion}} \\
\midrule
\textbf{R1} & The reasoning must make a single, unambiguous verdict and bind it to the analyzed defect site(s). \\
\textbf{R2} & The reasoning must not rely on unstated assumptions about specifications, libraries, or inputs. \\

\midrule
\multicolumn{2}{@{}l}{\textbf{Dimension: Consistency}} \\
\midrule
\textbf{R1} & The reasoning must not contain statements that contradict earlier claims. \\
\textbf{R2} & Terminology and abstraction levels must remain consistent throughout the explanation. \\
\textbf{R3} & The same code element must not be assigned conflicting roles. \\
\textbf{R4} & No conclusions may be drawn without sufficient grounding in the code. \\
\bottomrule
\end{tabularx}
\end{table}

\section{Additional Details of Limitation Analysis}
\label{app:limitation_analysis}

\lstdefinestyle{limitationcase}{
    basicstyle=\ttfamily\scriptsize,
    breaklines=true,
    breakatwhitespace=false,
    columns=fullflexible,
    keepspaces=true,
    showstringspaces=false,
    frame=single,
    framesep=3pt
}

\subsection{Representative Case Studies of Evaluator Limitations}
\label{app:limitation_case_studies}

we provide category-level comparative examples below. Each example is based on a real CodeRQ-Bench mismatch case and shows the evaluator limitation involved, how existing methods misjudge the reasoning, and how VERA addresses the issue.

\subsubsection{Evidence-Ungrounded Assessment}
\label{app:case:evidence_ungrounded}
\paragraph{Definition.}
The evaluator does not ground its assessment in the evidence needed to verify reasoning correctness in coding tasks, where many claims must be searched and checked against repository context, API documents, or task specifications.

\paragraph{Case.}
Case from CoderEval-RE, id 62b8b99de0d34b282c1811f8.
\begin{lstlisting}[style=limitationcase]
[TASK DESCRIPTION]
@classmethod
def _reset_logging(cls):
    """
    Reset
    """


[REASONING STEPS]
1. Import the logging module to access logging functionalities.
2. Call logging.shutdown() to flush and close all handlers.
3. Acquire the logging module's internal lock to ensure thread safety during modifications.
4. Retrieve the root logger instance.
5. Remove all existing handlers attached to the root logger.
6. Set the root logger's level back to the default WARNING level.
7. Iterate through all registered loggers in the manager's logger dictionary.
8. For each logger, remove all its handlers and reset its level to NOTSET.
9. Reset the logging.disable() settings to enable all log levels.
10. Release the logging module's internal lock to allow other threads to proceed.

[GENERATED ANSWER]
def _reset_logging(cls):
    import logging
    logging.shutdown()
    logging._acquireLock()
    try:
        root = logging.getLogger()
        for handler in root.handlers[:]:
            root.removeHandler(handler)
        root.setLevel(logging.WARNING)
        for logger_name in logging.root.manager.loggerDict:
            logger = logging.getLogger(logger_name)
            for handler in logger.handlers[:]:
                logger.removeHandler(handler)
            logger.setLevel(logging.NOTSET)
        logging.disable(logging.NOTSET)
    finally:
        logging._releaseLock()
\end{lstlisting}
\paragraph{Reasoning status.}
This is an incorrect reasoning chain because according to the Python logging documentation, \texttt{logging.shutdown()} is intended for application-exit shutdown after which the logging system should no longer be used; therefore, it should not be used as a runtime reset operation inside \texttt{\_reset\_logging}.

\paragraph{How existing methods overlook it.}
CaSE gives the chain the maximum score (3/3), marking all 10 steps as correct/relevant/coherent, and SocREVAL also assigns a high score (4/5). Both methods judge the reasoning mainly by local plausibility or agreement with a self-generated reference, so they do not verify whether \texttt{logging.shutdown()} satisfies the documented contract of Python logging. RECEVAL produces a borderline-low score ($\approx$ 0.50, just below the 0.5 threshold), but this comes from a single inter-step NLI contradiction false positive (min\_inter = 0.0012) rather than from identifying the actual issue.
\paragraph{How VERA identifies it.}
VERA invokes web search in Stage 1 and assigns a low score (\texttt{0.333}). It identifies the documented issue that \texttt{logging.shutdown()} is intended to close all handlers at application exit, after which no further use of the logging system should be made. Therefore, treating it as a runtime reset operation inside \texttt{\_reset\_logging} is not justified by the documented contract.

\subsubsection{Ambiguity Assessment Deficit}
\label{app:case:ambiguity_deficit}
\paragraph{Definition.}
The evaluator cannot determine whether a task is ambiguous or underspecified, nor assess whether the reasoning handles such ambiguity.

\paragraph{Case.}
Case from CoderEval-RE, id 62e4fc3c85ea98643089041e.
\begin{lstlisting}[style=limitationcase]
[TASK DESCRIPTION]
def _inline_r_setup(code: str) -> str:
    """
    Some behaviour of R cannot be configured via env variables, but can
    only be configured via R options once R has started. These are set here.
    """


[REASONING STEPS]
1. The function needs to prepend R configuration code to the provided input code.
2. Identify the R options that must be set at runtime. Common examples include:
   - Setting default CRAN mirror
   - Suppressing warnings
   - Disabling graphics
3. Format the R options string using valid R syntax
4. Concatenate the options string with the user-provided code
5. Return the combined string that will execute configuration first followed by user code

[GENERATED ANSWER]

def _inline_r_setup(code: str) -> str:
    r_options = (
        "options(repos=c(CRAN='https://cloud.r-project.org'), warn=-1, device=NULL)\n"
    )
    return r_options + code
\end{lstlisting}
\paragraph{Reasoning status.}
This reasoning chain is incorrect because the task states that some R options must be set at runtime, but does not specify which options or values. The reasoning confidently invents concrete options and hard-codes them without acknowledging that the task is underspecified or that repository/runtime context is needed.
\paragraph{How existing methods overlook it.}
CaSE assigns the maximum score (3/3) because the steps look coherent and plausible for “setting R options.” SocREVAL also gives a high score (4/5) because its self-generated reference is similarly generic and agrees with the candidate. Both methods miss the key issue: the reasoning is not merely filling in implementation details, but committing to unsupported choices under an underspecified prompt.
\paragraph{How VERA identifies it.}
VERA’s Stage 1 gives a borderline score (6/10, normalized 0.556), but Stage 2 detects task ambiguity (ambiguity\_level=0.5) and poor ambiguity handling (handling\_quality=0.2). The ambiguity-aware correction lowers the final score to 0.466, matching the human incorrect label. This shows how VERA penalizes confident commitment to arbitrary assumptions when the task specification is underspecified.

\subsubsection{Score Aggregation Failure}
\label{app:case:score_aggregation}
\paragraph{Definition.}
The evaluator produces an unreliable final score because its scoring rule combines step-level judgments in a way that either lets one misjudged step override an otherwise correct trace or allows fatal correctness errors to be diluted by other steps.

\paragraph{Case.}
Case from DebugBench-RE, id widest-vertical-area-between-two-points-containing-no-points\_cpp\_017\_correct.
\begin{lstlisting}[style=limitationcase]
[TASK DESCRIPTION]
Given n points on a 2D plane where points[i] = [xi, yi], Return the widest vertical area between two points such that no points are inside the area.
A vertical area is an area of fixed-width extending infinitely along the y-axis (i.e., infinite height). The widest vertical area is the one with the maximum width.
Note that points on the edge of a vertical area are not considered included in the area.

The following code attempts to solve the question above. Analyze whether it contains a bug.

class Solution {
public:
  int maxWidthOfVerticalArea(vector<vector<int>>& points) {
    set<int> x;
    int res = 0;
    for (auto &p : points)
        x.insert(p[0]);
    for (auto it = next(begin(x)); it != end(x); ++it)
        res = max(res, *it - *prev(it));
    return res;
}
};

[REASONING STEPS]
**Understanding the Problem:**
We are given `n` points on a 2D plane and tasked with finding the widest vertical area between any two points. The width is determined by the absolute difference in the x-coordinates of any two points.
**Input and Output:**
Input: A vector of vectors `points`, where each inner vector contains two integers representing the x and y coordinates.
Output: An integer representing the maximum width found between any two vertical lines formed by the x-coordinates of the points.
**Key Steps of the Code:**
**Collect Unique X-Coordinates:** The code uses a `set<int>` to store unique x-coordinates from the input points. This eliminates any duplicate x-coordinates automatically.
**Calculate Maximum Width:** By iterating through the sorted unique x-coordinates, the code computes the difference between consecutive elements to find the maximum width.
**Analysis of the Code:**
**Correctness of the Approach:**
Using a set ensures that the x-coordinates are unique and sorted.
Iterating through the sorted set and computing the difference between consecutive elements is a correct approach to finding the maximum distance between any two distinct x-coordinates.
**Edge Cases:**
If all points have the same x-coordinate, the set will have just one element, resulting in no valid vertical area between different x-coordinates, and `res` remains `0`, which is correct.
For more than one unique x-coordinate, the set will calculate and return the correct maximum difference.
**No Off-by-One Errors or Logical Errors:**
The code correctly uses `next(begin(x))` to start the comparison from the second element.
It uses `*it` and `*prev(it)` to calculate differences without any errors in accessing elements.

[GENERATED ANSWER]
1
\end{lstlisting}
\paragraph{Reasoning status.}
This reasoning chain is correct.
\paragraph{How existing methods misjudge it.}
This failure mainly appears in CaSE and RECEVAL, both of which rely on step-level scoring followed by hard aggregation. CaSE assigns a low score (\texttt{1/3}) even though 16 of 18 steps are marked fully correct/relevant/coherent; two non-content section headers receive local zeros, and CaSE’s AND-style aggregation propagates these local artifacts to the solution-level verdict. RECEVAL similarly assigns a low score (\texttt{$\approx$0.088}) because min-pooling lets a few brittle NLI/PMI bottlenecks dominate the final score.
\paragraph{How VERA identifies it.}
VERA avoids this failure by using a holistic chain-level score rather than step-level min/AND pooling. It assigns this correct reasoning chain a high score (\texttt{0.778}), matching the label.

\subsubsection{Self-Generated Reference Bias}
\label{app:case:self_generated_reference}
\paragraph{Definition.}
The evaluator judges reasoning against self-generated references or standards, creating circular validation in which shared errors or biases are mistaken for correctness.

\paragraph{Case.}
Case from CoderEval-RE, id 6306092d73426c38ae68ad06
\begin{lstlisting}[style=limitationcase]
[TASK DESCRIPTION]
    def get_silent_args(self, args):
        """list of silenced argument

        :param args: The received arguments.
        :return: list, slienced argument names
        """


[REASONING STEPS]
1. Convert the input `args` (assumed to be an argparse.Namespace object) into a dictionary using `vars(args)`.
2. Iterate through each argument-value pair in the dictionary.
3. Check if the value of the argument is `False`.
4. Collect the names of all arguments where the value is `False`.
5. Return the collected list of silenced argument names.

[GENERATED ANSWER]

def get_silent_args(self, args):
    return [arg for arg, val in vars(args).items() if val is False]
\end{lstlisting}
\paragraph{Reasoning status.}
This reasoning chain is incorrect. The prompt asks for “silenced” argument names, but the reasoning invents the rule that silenced $\Leftrightarrow$ value is False. In argparse, a False value can simply come from a default value or an unpassed flag, and does not necessarily indicate that an argument is “silenced.”
\paragraph{How existing methods misjudge it.}
This failure is most clearly shown by SocREVAL, which assigns the maximum score (\texttt{5/5}) because its self-generated reference makes the same assumption as the candidate: convert \texttt{args} to a dictionary, filter values that are \texttt{False}, and return the keys. As a result, SocREVAL rewards the candidate for matching its own generated answer, even though both rely on the same unsupported interpretation of “silenced.”
\paragraph{How VERA identifies it.}
VERA avoids this self-confirmation loop by not relying on a self-generated reference. It assigns a low score (0.333) and identifies that False values depend on argument definitions and defaults, so they are not a reliable criterion for identifying “silenced” arguments.

\subsubsection{Code-Unaware Assessment}
\label{app:case:code_unaware}
\paragraph{Definition.}
The evaluator fails to reliably assess code-related reasoning because it cannot correctly represent or judge code-relevant semantics in the reasoning process.

\paragraph{Case.}
Case from CoderEval-RE, id 62e4fb6585ea98643089032b.
\begin{lstlisting}[style=limitationcase]
[TASK DESCRIPTION]
def parse_version(s: str) -> tuple[int, ...]:
    """poor man's version comparison"""


[REASONING STEPS]
1. Split the input string using the dot ('.') as the delimiter. This creates a list of string components.
2. Convert each component from a string to an integer using a mapping function.
3. Convert the resulting list of integers into a tuple.
4. Return the tuple of integers.

[GENERATED ANSWER]

def parse_version(s: str) -> tuple[int, ...]:
    return tuple(map(int, s.split('.')))
\end{lstlisting}
\paragraph{Reasoning status.}
This reasoning chain is correct.
\paragraph{How existing methods overlook it.}
This failure mainly appears in RECEVAL, which assigns a score (0.21). RECEVAL decomposes the reasoning into premise-conclusion pairs with SRL and then applies an NLI model trained for natural-language entailment. As a result, it fails to recognize code-specific facts such as str.split(".") returning a list of string components, and it treats sequential imperative steps such as “split,” “convert,” and “return” as possible contradictions. The chain is correct, but the code-unaware SRL/NLI pipeline misrepresents its semantics.
\paragraph{How VERA identifies it.}
VERA uses a code-capable LLM judge over the full reasoning chain and generated code, rather than converting code reasoning into brittle natural-language entailment pairs. It assigns the chain the maximum score (1.000), correctly recognizing the Python semantics and matching the label.

For each case study above, we base the analysis on the evaluator scores, intermediate outputs, and diagnostic annotations for that specific instance.

\subsection{Implementation Details of the Limitation Analysis Pipeline}
\label{app:implementation_details}

\subsubsection{Diagnostic Annotation}
\label{app:diagnostic_annotation}

We use GPT-5.2 to generate diagnostic annotations for mismatch cases.
The model is provided with the original task, the LLM-generated reasoning chain and final output, and the evaluator's detailed assessment outputs for that reasoning.
The diagnostic annotation stage uses two prompt variants corresponding to the two mismatch types defined in \S~\ref{sec:analysis_setup_and_protocol}: \textit{missed errors} and \textit{false alarms}.

\paragraph{Prompt for Missed Errors}
\label{app:prompt_diagnostic_missed}

This prompt is used for mismatch cases in which the evaluator assigns a high score to reasoning that is actually incorrect. The full prompt is shown in Table~\ref{tab:prompt_missed_error}.

\begin{table*}[h]
  \centering
  \begin{aclprompt}{PROMPT TEMPLATE: DIAGNOSTIC ANNOTATION FOR MISSED ERRORS}
    You are an expert analyzing why the reasoning evaluation method \var{method\_name} produces scores/conclusions that significantly differ from human evaluation when assessing LLM-generated reasoning. \\
    \\
    \textbf{\#\# Background} \\
    This is a case where \var{method\_name} gave a high score, but the reasoning is actually incorrect (human label = 0). Your task is NOT to judge whether the LLM's reasoning is correct (this is already determined by human annotation). Instead, analyze what content the evaluation method \var{method\_name} failed to assess in this LLM-generated reasoning, causing it to give a high score while human annotators marked it as incorrect. \\
    \\
    \textbf{\#\# \var{method\_name} Evaluation Method Details} \\
    \var{method\_description} \\
    \\
    \textbf{\#\# Case Content} \\
    \var{case\_content} \\
    \\
    \textbf{\#\# Your Analysis Task} \\
    The "Execution Log" section above shows the actual intermediate outputs from the \var{method\_name} pipeline for this case. Use this evidence to pinpoint exactly where and why the evaluation failed. Specifically: \\
    1. Identify which specific component/step in the pipeline produced the erroneous signal (e.g., which SRL parse was wrong, which NLI judgment was incorrect, which dimension scored too high, or how the self-generated reference was flawed). \\
    2. Explain the root cause: why did that component fail on this particular input? \\
    3. What capability is missing from \var{method\_name} that would be needed to correctly evaluate this case?
  \end{aclprompt}
  \caption{Prompt template used to generate diagnostic annotations for missed-error cases, where the evaluator assigns a high score to reasoning that is actually incorrect. Here, \var{case\_content} includes the original task, the LLM-generated reasoning chain and final output, and, the execution log containing the evaluator's intermediate assessment outputs. Text in blue braces denotes dynamic variables.}
  \label{tab:prompt_missed_error}
\end{table*}

\paragraph{Prompt for False Alarms}
\label{app:prompt_diagnostic_false_alarm}

This prompt is used for mismatch cases in which the evaluator assigns a low score to reasoning that is actually correct. The full prompt is shown in Table~\ref{tab:prompt_false_alarm}.

\begin{table*}[h]
  \centering
  \begin{aclprompt}{PROMPT TEMPLATE: DIAGNOSTIC ANNOTATION FOR FALSE ALARMS}
    You are an expert analyzing why the reasoning evaluation method \var{method\_name} produces scores/conclusions that significantly differ from human evaluation when assessing LLM-generated reasoning. \\
    \\
    \textbf{\#\# Background} \\
    This is a case where \var{method\_name} gave a low score, but the reasoning is actually correct (human label = 1). Your task is NOT to judge whether the LLM's reasoning is correct (this is already determined by human annotation). Instead, analyze what content the evaluation method \var{method\_name} may have incorrectly penalized in this LLM-generated reasoning, causing it to give a low score while human annotators marked it as correct. \\
    \\
    \textbf{\#\# \var{method\_name} Evaluation Method Details} \\
    \var{method\_description} \\
    \\
    \textbf{\#\# Case Content} \\
    \var{case\_content} \\
    \\
    \textbf{\#\# Your Analysis Task} \\
    The "Execution Log" section above shows the actual intermediate outputs from the \var{method\_name} pipeline for this case. Use this evidence to pinpoint exactly where and why the evaluation failed. Specifically: \\
    1. Identify which specific component/step in the pipeline produced the erroneous signal (e.g., which step was the bottleneck, which SRL parse was malformed, which NLI score was unreasonably low, or which dimension was incorrectly penalized). \\
    2. Explain the root cause: why did that component fail on this particular input? \\
    3. What are the characteristics of this correct reasoning that confused the \var{method\_name} evaluator?
  \end{aclprompt}
  \caption{Prompt template used to generate diagnostic annotations for false-alarm cases, where the evaluator assigns a low score to reasoning that is actually correct. Here, \var{case\_content} includes the original task, the LLM-generated reasoning chain and final output, and, when available, the execution log containing the evaluator's intermediate assessment outputs. Text in blue braces denotes dynamic variables.}
  \label{tab:prompt_false_alarm}
\end{table*}

\subsubsection{Open-Coded Limitation Label Extraction}
\label{app:label_extraction}

In the second stage of the pipeline, we use GPT-5.2 to extract open-coded limitation labels from the diagnostic annotations generated in the previous stage.
These labels are intended to capture the specific failure modes reflected in each mismatch case before taxonomy consolidation into higher-level categories.
The full prompt is shown in Table~\ref{tab:prompt_open_coding}.

\begin{table*}[t]
  \centering
  \begin{aclprompt}{PROMPT TEMPLATE: OPEN-CODED LIMITATION LABEL EXTRACTION}
    You are analyzing a case in which \var{method\_name}, an evaluation method for assessing LLM-generated reasoning on coding tasks, produced an assessment that is misaligned with the actual correctness of the reasoning. \\
    \\
    \textbf{\#\# Mismatch Type: \var{mismatch\_type}} \\
    \var{mismatch\_description} \\
    \\
    \textbf{\#\# Diagnostic Annotation} \\
    The following diagnostic annotation explains why \var{method\_name} failed on this case: \\
    \var{annotation} \\
    \\
    \textbf{\#\# Task} \\
    Based on the diagnostic annotation above, extract 2--4 \textbf{limitation labels} that describe why \var{method\_name} failed in this case. \\
    \\
    Key question: What limitation caused \var{method\_name} to make this misjudgment? \\
    \\
    \textbf{\#\# Requirements} \\
    - Derive the labels directly from the diagnostic annotation \\
    - Each label should contain 2--5 words, written in lowercase with underscores \\
    - Focus on limitations in evaluating LLM-generated reasoning for coding tasks \\
    - Do not use overly generic labels that could apply to any evaluation error \\
    \\
    Return JSON: \\
    \{\{"limitations": ["label1", "label2", ...]\}\}
  \end{aclprompt}
  \caption{Unified prompt template used to extract open-coded limitation labels from diagnostic annotations in the second stage of the pipeline. Here, \var{method\_name} denotes the evaluation method being analyzed; \var{mismatch\_type} denotes the mismatch type of the case, whose value is either \textit{missed errors} or \textit{false alarms}; \var{mismatch\_description} provides the corresponding textual description of that mismatch type; and \var{annotation} denotes the diagnostic annotation generated in the previous stage for the case. The prompt returns 2--4 limitation labels in JSON format. Text in blue braces denotes dynamic variables.}
  \label{tab:prompt_open_coding}
\end{table*}

\subsubsection{Taxonomy Consolidation}
\label{app:taxonomy_consolidation}

In the third stage of the pipeline, we consolidate the open-coded limitation labels into higher-level limitation categories.
This stage follows a two-step process: embedding-based pre-clustering using \texttt{text-embedding-3-large}, followed by taxonomy consolidation using GPT-5.2.
The pre-clustering step groups semantically similar labels before GPT-based consolidation into the final taxonomy.
The full prompt for taxonomy consolidation is shown in Table~\ref{tab:prompt_taxonomy_consolidation}.

\begin{table*}[t]
  \centering
  \begin{aclprompt}{PROMPT TEMPLATE: TAXONOMY CONSOLIDATION}
    You are given \var{num\_pcs} label clusters (called "preclusters") that describe mismatch patterns of automated methods for evaluating Chain-of-Thought reasoning in coding tasks. \\
    \\
    These preclusters were extracted from 1,069 mismatch cases across three coding tasks (code generation, code summarization, bug detection) and three evaluation methods (CaSE, ReCEval, SocREval). Each precluster name describes a specific way an evaluation method mismatches---either by missing an error in reasoning or by falsely flagging correct reasoning. \\
    \\
    \textbf{\#\# All \var{num\_pcs} Preclusters (sorted by frequency):} \\
    \var{formatted\_preclusters} \\
    \\
    \textbf{\#\# Your Task} \\
    Group these \var{num\_pcs} preclusters into categories that are strictly MECE (Mutually Exclusive and Collectively Exhaustive). \\
    \\
    \textbf{\#\#\# MECE Requirements} \\
    \textbf{Mutually Exclusive} means: \\
    - For ANY given precluster, there must be exactly ONE category it belongs to---no ambiguity. \\
    - Two categories X and Y are mutually exclusive if and only if: there is NO conceivable mismatch instance that could be equally well described by both X and Y. \\
    - Conceptual test: If someone describes a mismatch and you cannot decide whether it is X or Y, then X and Y are NOT mutually exclusive and must be merged or re-divided along a clearer fault line. \\
    \\
    \textbf{Collectively Exhaustive} means: \\
    - Every single one of the \var{num\_pcs} preclusters must fit into exactly one category. \\
    - No residual "other" or "miscellaneous" category. \\
    \\
    \textbf{\#\#\# What NOT to do} \\
    - Do NOT create categories based on evaluation method names (CaSE, ReCEval, SocREval)---categories should describe the nature of the mismatch, not which tool has it. \\
    - Do NOT create categories that are subsets of other categories. \\
    - Do NOT create categories that have causal relationships (e.g., "shallow analysis" causing "logic errors")---if A commonly causes B, they are not independent dimensions and need rethinking. \\
    \\
    \textbf{\#\#\# Instructions} \\
    1. First, read through all \var{num\_pcs} preclusters and identify the natural fault lines. \\
    2. Propose categories. For each category, provide: \\
    \quad - A clear \textbf{name} (2--5 words, lowercase\_with\_underscores) \\
    \quad - A precise \textbf{definition} (1--2 sentences) \\
    \quad - A \textbf{boundary rule}: how to decide if a precluster belongs here vs. in another category \\
    3. After proposing all categories, run a \textbf{pairwise MECE check}: for each pair (X, Y), state in one sentence why a mismatch described by X cannot also be described by Y. If you find overlap, revise. \\
    4. Finally, assign every precluster to exactly one category. \\
    \\
    \textbf{\#\#\# Output Format} \\
    Return a JSON object: \\
    \texttt{\{ "reasoning": "Your analysis of the natural fault lines you identified",} \\
    \quad \texttt{"categories": [ \{} \\
    \quad \quad \texttt{"name": "category\_name", "definition": "Precise definition of this category",} \\
    \quad \quad \texttt{"boundary\_rule": "When a precluster is X vs Y, it belongs here if...",} \\
    \quad \quad \texttt{"assigned\_preclusters": ["precluster1", "precluster2", ...] \} ],} \\
    \quad \texttt{"pairwise\_checks": [ \{ "pair": ["category\_a", "category\_b"],} \\
    \quad \quad \texttt{"why\_exclusive": "One sentence explaining mutual exclusivity" \} ] \}}
  \end{aclprompt}
  \caption{Prompt template used for GPT-based taxonomy consolidation in the third stage of the pipeline. Here, \var{num\_pcs} denotes the number of preclusters produced after embedding-based pre-clustering, and \var{formatted\_preclusters} denotes the formatted list of those preclusters together with their frequency, size, and task-distribution information. The prompt asks the model to consolidate these preclusters into a final taxonomy and return category names, definitions, boundary rules, assigned preclusters, and consistency checks in JSON format. Text in blue braces denotes dynamic variables.}
  \label{tab:prompt_taxonomy_consolidation}
\end{table*}

\section{Additional Details of \methodname{}}

\subsection{Prompt for Verified Reasoning Evaluation}
\label{app:verified_prompt}

Table~\ref{tab:verified_reasoning_prompt} presents the prompt template used in the first stage of \methodname{}.
Given a coding-task description, an LLM-generated reasoning chain, and the corresponding output, this prompt instructs the evaluator to assess the overall correctness of the reasoning process.
When technical claims require external verification, the evaluator is allowed to consult web search for evidence grounding.
The returned score is an integer from 1 to 10, which is later normalized to $[0,1]$ as described in \S~\ref{sec:method}.

\begin{table*}[h]
  \centering
  \begin{aclprompt}{Prompt for Verified Reasoning Evaluation}
    You are a strict evaluator of AI reasoning chains in coding tasks, with access to web search. Most AI-generated reasoning contains errors --- your job is to find them. \\
    \\
    \textbf{\#\# Task / Question:} \\
    \{question\} \\
    \\
    \textbf{\#\# AI's Reasoning Chain:} \\
    \{chain\} \\
    \\
    \textbf{\#\# AI's Generated Output:} \\
    \{output\} \\
    \\
    Use web search to verify specific technical claims when needed (API usage, library functions, language features, algorithm correctness). \\
    \\
    Evaluate the CORRECTNESS of the reasoning chain and whether it leads to a correct solution: \\
    - Does the reasoning correctly identify the key technical aspects of the task? \\
    - Are claims about algorithms, data structures, edge cases, and syntax correct? \\
    - Does the reasoning lead to an output that correctly solves the problem? \\
    - Are there factual errors about language features, APIs, or libraries in the reasoning? \\
    \\
    Rate the reasoning correctness from 1 to 10 (most samples should fall in 3--7 range; reserve 9--10 only for truly flawless reasoning): \\
    10 = Flawless reasoning; all claims correct, leads to correct solution \\
    8--9 = Strong; minor issues that do not affect the solution's correctness \\
    6--7 = Acceptable; one moderate technical error or oversight in reasoning \\
    4--5 = Problematic; significant error that would affect solution correctness \\
    2--3 = Poor; multiple reasoning errors that would produce wrong results \\
    1 = Fundamentally wrong; core reasoning approach is incorrect \\
    \\
    Respond with ONLY a JSON object: \\
    \{"score": <1-10>, "reason": "brief justification"\}
  \end{aclprompt}
  \caption{Prompt template used in the verified reasoning evaluation stage of \methodname{}. The placeholders \{question\}, \{chain\}, and \{output\} are instantiated with the task description, the candidate reasoning chain, and the corresponding generated output, respectively.}
  \label{tab:verified_reasoning_prompt}
\end{table*}

\subsection{Prompt for Ambiguity-Aware Score Correction}
\label{app:ambiguity_prompt}

Table~\ref{tab:ambiguity_correction_prompt} presents the prompt template used in the second stage of \methodname{}.
Given the task description, the reasoning chain, and the generated output, the evaluator estimates two quantities: the ambiguity level of the task and the quality of the model's uncertainty handling.
These two scores are then used to compute the ambiguity-aware correction term described in \S~\ref{sec:method}.

\begin{table*}[h]
  \centering
  \begin{aclprompt}{Prompt for Ambiguity-Aware Score Correction}
    You are evaluating how well an AI's reasoning chain handles uncertainty in a coding task. \\
    \\
    \textbf{\#\# Task / Question:} \\
    \{question\} \\
    \\
    \textbf{\#\# AI's Reasoning Chain:} \\
    \{chain\} \\
    \\
    \textbf{\#\# AI's Generated Output:} \\
    \{output\} \\
    \\
    Assess two things: \\
    \\
    \textbf{\#\#\# 1. Task Ambiguity Level (0.0 - 1.0)} \\
    How ambiguous or underspecified is this task for a software developer? \\
    \\
    Use this calibrated scale --- note that scores above 0.5 should be common for real-world tasks: \\
    - 0.9--1.0 = Fundamentally underspecified: no single correct answer; any experienced developer would need to ask clarifying questions before starting. E.g., ``improve this function,'' ``fix the bug,'' open-ended refactoring, or missing success criteria. \\
    - 0.6--0.8 = Significantly ambiguous: the task has a plausible main interpretation, but key design decisions (error handling, edge cases, performance/readability trade-offs, API style) are left unstated and reasonable developers would make different choices. \\
    - 0.3--0.5 = Mildly ambiguous: mostly clear, but one or two specific behaviors are underspecified (e.g., what to return on empty input, whether to mutate or copy, which exception type to raise). \\
    - 0.1--0.2 = Essentially clear: a well-specified algorithmic task; at most trivial naming choices remain. \\
    - 0.0 = Fully deterministic: one unambiguous correct answer, no design decisions required. \\
    \\
    \textbf{\#\#\# 2. Uncertainty Handling Quality (0.0 - 1.0)} \\
    Only matters when ambiguity\_level > 0.2. How appropriately does the AI's reasoning respond to the actual ambiguity? \\
    \\
    Strict scale --- most reasoning chains will score low because they ignore ambiguity: \\
    - 0.8--1.0 = Excellent: explicitly names the ambiguity, states which interpretation was chosen and why, preserves conditional conclusions (``if X, then A; otherwise B''), or reasonably abstains on genuinely unknowable aspects. \\
    - 0.5--0.7 = Partial: acknowledges uncertainty somewhere in the reasoning but does not fully address it (e.g., mentions ``assuming X'' once without justification). \\
    - 0.2--0.4 = Poor: proceeds confidently with one interpretation, makes large implicit assumptions, and gives no acknowledgment of alternatives. \\
    - 0.0--0.1 = Actively misleading: presents a highly contested design choice as the only correct approach, or expresses false certainty about unknowable behavior. \\
    \\
    Default: if ambiguity\_level < 0.2, set handling\_quality = 0.5 (neutral --- the task is clear, so uncertainty handling is not relevant). \\
    \\
    Respond with ONLY a JSON object: \\
    \{"ambiguity\_level": <0.0-1.0>, "handling\_quality": <0.0-1.0>, "handling\_issues": "brief explanation"\}
  \end{aclprompt}
  \caption{Prompt template used in the ambiguity-aware score correction stage of \methodname{}. The placeholders \{question\}, \{chain\}, and \{output\} are instantiated with the task description, the candidate reasoning chain, and the corresponding generated output, respectively.}
  \label{tab:ambiguity_correction_prompt}
\end{table*}

\section{Additional Details of Experiment}
\label{app:experiment}

\subsection{Baseline Methods}
\label{app:baseline_overview}

\paragraph{RECEVAL}
RECEVAL (Reasoning Chain Evaluation)~\cite{prasad2023receval} is a reference-free framework for evaluating multi-step reasoning chains, originally developed for tasks such as deductive reasoning, mathematical word problems, and discrete reading comprehension. It conceptualizes a reasoning chain as an informal proof and decomposes each step into fine-grained Reasoning Content Units (RCUs) using semantic role labeling. The method evaluates reasoning along two primary dimensions, correctness and informativeness: correctness is measured through both intra-step and inter-step logical consistency using natural language inference, while informativeness is quantified by pointwise V-information to capture each step’s information gain toward the final answer. In its original form, RECEVAL outputs continuous chain-level scores for these dimensions by applying minimum-pooling aggregation over step-level evaluations, where correctness is a probability in $(0,1)$ and informativeness is an unbounded real-valued score.

\paragraph{SOCREVAL}
SOCREVAL (Socratic Method-Inspired Reasoning Evaluation)~\cite{he2024socreval} is a reference-free framework for assessing machine-generated reasoning chains without model fine-tuning or human-written references. It uses large language models prompted with Socratic principles, including Definition, Maieutics, and Dialectic, to guide qualitative evaluation of reasoning. The evaluator first generates its own response to the input prompt, then analyzes the candidate reasoning chain, and finally assigns an overall reasoning-quality judgment. The original method outputs a discrete score from 1 to 5 and was developed and validated on a range of general reasoning tasks, including arithmetic reasoning, deductive and commonsense inference, and discrete reasoning over paragraphs.

\paragraph{CaSE}
CaSE (Causal Stepwise Evaluation)~\cite{do2025defines} is a step-level framework for reasoning evaluation, originally developed for mathematical problem-solving tasks. To reduce hindsight bias and better reflect autoregressive reasoning generation, it evaluates each intermediate step using only the original question and the preceding context available before that step. The method assesses reasoning mainly along two dimensions, relevance and coherence, where relevance measures whether a step is grounded in the problem and coherence measures whether it logically follows from prior steps. In its original form, CaSE outputs explicit binary judgments (0 or 1) for each reasoning step on these dimensions, providing localized fine-grained evaluations rather than a single holistic score.

\paragraph{MAD}
MAD (Multi-Agent Debate) is adapted by Zhang et al.~\cite{zhang2025they} for detecting low-quality reasoning through structured multi-agent debate. In this setup, a verifier raises potential logical flaws in a candidate reasoning chain, a defender responds with counterarguments and supporting evidence, and an arbiter evaluates the rationality of the debate outcome. The original method produces a final binary decision after a fixed number of debate rounds. 

\paragraph{LLM-as-Judge}
LLM-as-Judge~\cite{zheng2023judging} is a general evaluation paradigm that uses strong large language models to assess the quality of candidate responses across diverse tasks, including coding and mathematical reasoning. It supports both pairwise comparison and direct scoring; in the latter setting, the judge assigns a score, typically with a textual explanation. In our experiments, we use it as a vanilla LLM-based evaluator for overall reasoning quality assessment.

\paragraph{AutoRace}
AutoRace~\cite{hao2024new} is a fully automated framework for evaluating step-by-step reasoning chains, originally developed for mathematical, commonsense, and logical reasoning tasks. It operates by first inducing task-specific evaluation criteria from collected incorrect reasoning chains, using a LLM to identify and summarize common error patterns. The resulting criteria are then used to guide detailed evaluation of a candidate reasoning chain. Since these criteria are generated dynamically for each task, the evaluation dimensions are task-adaptive rather than fixed, and the original method outputs a binary judgment for the overall reasoning process.

\subsection{Evaluation Metrics}
\label{app:metrics}
Somers'~D and Spearman's~$\rho$, both defined in $[-1,1]$, measure the ordinal and monotonic association between predicted scores and ground-truth correctness labels, respectively. AUCROC measures the probability that a correct reasoning trace receives a higher score than an incorrect one, while AUPRC is particularly informative under class imbalance.

\subsection{Implementation Details}
\label{app:implementation}

We follow the original configurations of all baselines whenever possible. RECEVAL retains its original correctness scores, with informativeness normalized to $[0,1]$ before averaging the three reported dimensions into a unified score. SOCREVAL and CaSE use GPT-4 by default and GPT-4.1 on SWEbench-RE due to context length limits.

VERA, MAD$^{*}$, LLM-as-Judge, and AutoRace$^{*}$ are all instantiated with GPT-4o-mini for fair comparison. Each MAD debate runs for three rounds. For AutoRace$^{*}$, we first collect incorrect reasoning examples using Qwen-2.5 with manual verification, and then use GPT-4o-mini for both criteria induction and final evaluation.

VERA uses $\tau=0.4$, selected based on the threshold sensitivity study in Appx.~\ref{app:tau_ablation}. All baseline outputs are normalized to $[0,1]$ for comparability.

\begin{figure}[!t]
    \centering
    \includegraphics[width=0.9\linewidth]{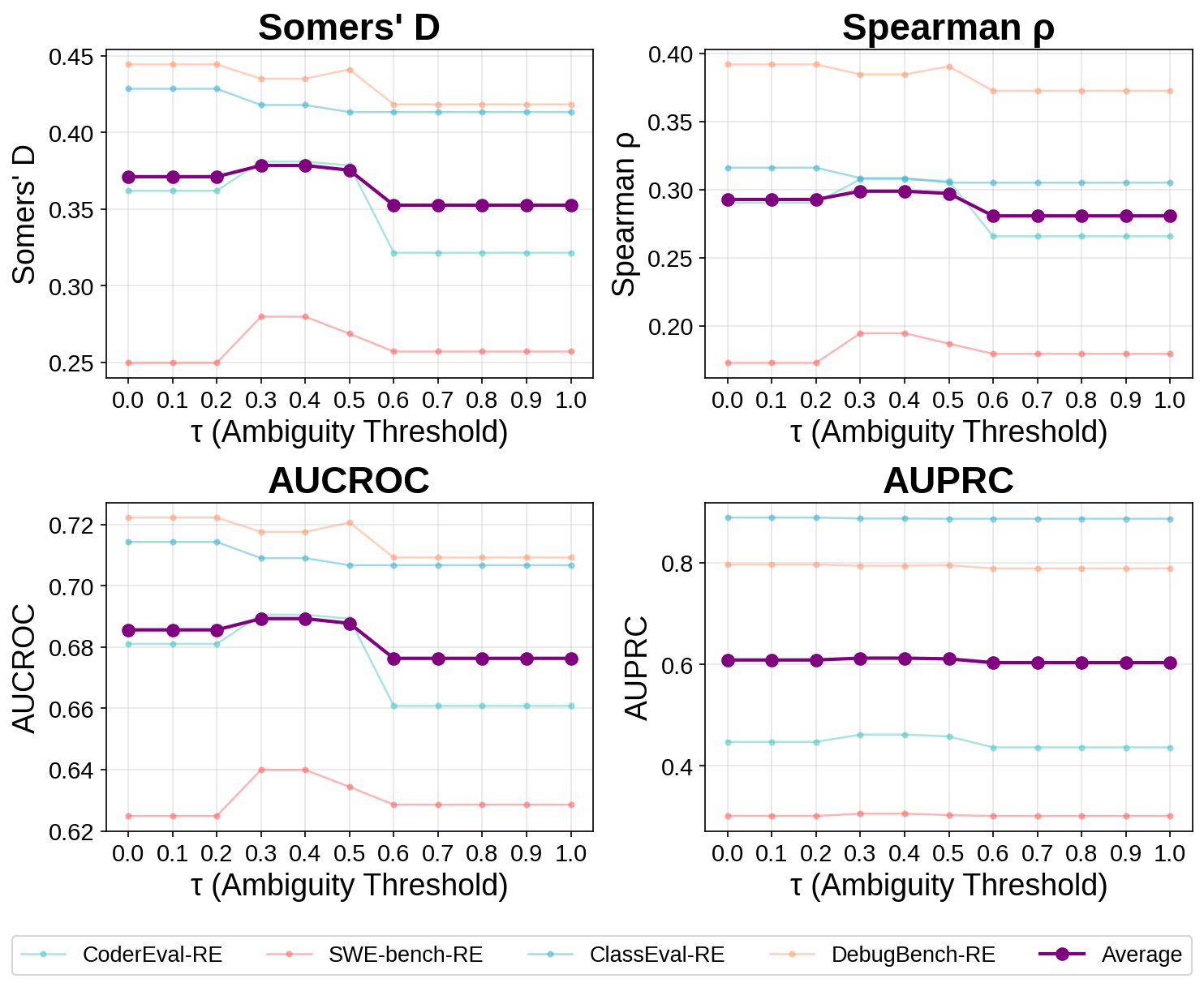}
    \caption{Threshold sensitivity analysis for the ambiguity threshold $\tau$ in \methodname{}.}
    \label{fig:tau_ablation}
\end{figure}

\subsection{Threshold Sensitivity of the Ambiguity Correction}
\label{app:tau_ablation}

We analyze the sensitivity of \methodname{} to the ambiguity threshold $\tau$ used in the ambiguity-aware correction module. To do so, we vary $\tau$ from $0.0$ to $1.0$ and evaluate the resulting performance on all four datasets using the same metrics as in the main experiments.

Fig.~\ref{fig:tau_ablation} summarizes the results. Overall, \methodname{} achieves the strongest average performance around $\tau=0.4$, which we therefore use in all experiments. Smaller values make ambiguity-aware correction overly aggressive, whereas larger values make it too conservative. The effect is more pronounced on the generation-oriented datasets, while the summarization and bug detection datasets are relatively less sensitive to the threshold.

\end{document}